\documentstyle[12pt,a4]{article}
\begin{document}
\begin{titlepage}
\begin{flushright}
ECT*-96-010\\
TK 96 18\\
NORDITA 96/35 N,P\\[1cm]
\end{flushright}
\vfill
\begin{center}
{\bf Workshop on}\\[1cm]
{\Huge THE STANDARD MODEL\\[1cm] AT LOW ENERGIES}\\[1cm]
ECT*, Trento, Italy, April 29 --- May 10, 1996\\
\vfill
{\large ABSTRACT BOOKLET}\\[1cm]
\end{center}
These are short proceedings of the workshop on ``The Standard Model at
Low Energies'' held at ECT* in Trento, Italy, from April 29 to May 10, 1996.
The workshop concentrated on Chiral Perturbation Theory in
its various settings.
Included are a one page abstract with references per speaker and
a listing of some review papers of relevance to the field.\\[2cm]
\begin{center}
{\bf J.~Bijnens}\\NORDITA, Blegdamsvej 17,\\
DK-2100 Copenhagen \O, Denmark\\[0.5cm]
{\bf U.-G.~Mei{\ss}ner}\\
Universit\"at Bonn, Institut f\"ur Theoretische Kernphysik, Nussallee 14-16,\\
D-53115 Bonn, Germany
\end{center}
\end{titlepage}
\section{Introduction}
The field of Chiral Perturbation Theory is a very active one.
It was therefore felt that another topical workshop was needed.
This meeting followed
the series of workshops in Ringberg (Germany), 1988, Dobog/'ok/"o
(Hungary), 1991, and Karreb\ae{}ksminde (Denmark), 1993,
which were all three of the size of about 50 participants.
We were lucky enough to obtain funding from the European Center for
Theoretical Studies in Nuclear Physics and related Areas, the ECT*,
 for this workshop.

The meeting itself took place in Trento in the spring of 1996. There were
a lot of discussions and talks. The present abstract booklet is meant as a
guide into the literature and provides abstracts and main references of the
presentations given. Since most results will be published elsewhere,
a full traditional type of proceedings seemed unnecessary.

We would like to take the opportunity to thank the ECT*, its director, Ben
Mottelson, and its board of directors for general support and enthusiasm
for this meeting. We would also like to thank the secretaries Cristina Costa
at ECT* and Anne Lumholdt at NORDITA for taking care of the administrative
work involved. Finally we thank the participants for making this a very
pleasant and lively meeting.

Unfortunately, we cannot thank the weatherman since it rained most of the
workshop, but this way the presence of most participants at the talks was
guaranteed.

There exists by now a rather large series of main references related
to Chiral Perturbation Theory. We give below a short and subjective
 list of the most recent
ones and those considered classics. Then the list of participants and their
electronic mail addresses follows. The program of the meeting and
the individual
abstracts with their references round off this abstract booklet. The latter
are given in the same order as the talks were given. Speakers are indicated
in boldface.

\begin{flushright}
Johan Bijnens and Ulf-G. Mei{\ss}ner
\end{flushright}
\newpage
\section{A short guide to review literature}
Chiral Perturbation Theory grew out of current algebra, and it  soon
was realized that certain terms beyond the lowest order were also uniquely
defined. This early work and references to earlier review papers can be found
in [1]. Weinberg then proposed a systematic method in [2], later systematized
and extended to use the external field method in the classic papers
by Gasser and Leutwyler [3,4], which, according to Howard Georgi,
everybody should put under his/her pillow before he/she goes to sleep.
The field has since then extended a lot
and relatively recent review papers are: Ref.[5] with an emphasis on the
anomalous sector, Ref.[6] giving a general overview over the vast field
of applications in  various areas of physics, Ref.[7] on mesons and baryons,
and Ref.[8] on baryons and multibaryon processes. In addition there are books
by Georgi[9], which, however, does not cover the standard approach,
including the terms in the lagrangian at higher order and
a more recent one by Donoghue, Golowich and Holstein[10].

There are also the lectures available on the archives by E.~de~Rafael [11]
and A.~Pich[12] as well as numerous others. The references to the previous
meetings are [13,14,15]. There are also the MIT meeting [16] and the DA$\Phi$NE
handbook [17] as useful references.

{\bf References}\\
{}[1] H. Pagels, Phys. Rep. 16 (1975) 219\\
{}[2] S. Weinberg, Physica 96A (1979) 327\\
{}[3] J. Gasser and H. Leutwyler, Ann. Phys. (NY) 158 (1984) 142\\
{}[4] J. Gasser and H. Leutwyler, Nucl. Phys. B250 (1985) 465\\
{}[5] J. Bijnens, Int. J. Mod. Phys. A8 (1993) 3045\\
{}[6] U.-G. Mei{\ss}ner, Rep. Prog. Phys. 56 (1993) 903\\
{}[7] G. Ecker, Prog. Nucl. Part. Phys. 35 (1995) 1\\
{}[8] V. Bernard, N. Kaiser and U.-G. Mei{\ss}ner,
Int. J. Mod. Phys. E4 (1995) 193\\
{}[9] H.~Georgi, {\it Weak Interactions and Modern Particle Theory}, 1984,
Benjamin/Cummings.\\
{}[10] J.~Donoghue, E. Golowich and B. Holstein, {\it Dynamics of the Standard
Model}, Cambridge University Press.\\
{}[11] E. de Rafael, hep-ph/9502254, lectures at the TASI-94 Summer
  School, ed. J.F. Donoghue, World Scientific, Singapore, 1995.\\
{}[12] A.~Pich, Rep. Prog. Phys. 58 (1995) 563, hep-ph 9505231\\
{}[13] A. Buras, J.-M. G\'erard and W. Huber (eds.),
Nucl. Phys. B (Proc. Suppl) 7A (1989)\\
{}[14] U.-G.~Mei{\ss}ner (ed.), {\it Effective Field Theories of the
  Standard Model}, World Scientific, Singapore, 1992 \\
{}[15] J.~Bijnens, NORDITA-93/73.\\
{}[16] A.~Bernstein and B.~Holstein (eds.), {\it Chiral Dynamics : Theory and
 Experiment}, Springer Verlag, 1995.\\
{}[17] L. Maiani, G. Pancheri and N. Paver (eds.), {\it The
Second DA$\Phi$NE Physics Handbook}, 1995, SIS-Pubblicazione dei Laboratori
Nazionali di Frascati, P.O.Box 13, I-00044 Frascati, Italy.
\newpage
\section{Participants and their email}
\begin{verbatim}
S.Beane         sbeane@phy.duke.edu
V.Bernard       bernard@crnhp4.in2p3.fr
J.Bijnens       bijnens@nordita.dk
B.Borasoy       borasoy@pythia.itkp.uni-bonn.de
M.Butler        mbutler@ap.stmarys.ca
P.Buettiker     buettike@butp.unibe.ch
G.Colangelo     colangel@butp.unibe.ch
F.Cornet        cornet@ugr.es
G.D'Ambrosio    dambrosio@axpna1.na.infn.it
D.Drechsel      drechsel@vkpmza.kph.uni-mainz.de
S.Duerr         stephan@sonne.physik.unizh.ch
G.Ecker         ecker@ariel.pap.univie.ac.at
H.Forkel        forkel@ect.unitn.it
A.Gall          gall@butp.unibe.ch
J.Gasser        gasser@butp.unibe.ch
E.Golowich      golowich@phast.umass.edu
C.Hanhart       kph168@ikp187.ikp.kfa-juelich.de
T.Hannah        hannah@frodo.mi.aau.dk
T.Hemmert       hemmert@phast.umass.edu
G.Isidori       isidori@vaxlnf.lnf.infn.it
N.Kaiser        nkaiser@physik.tu-muenchen.de
J.Kambor        kambor@ipncls.in2p3.fr
M.Knecht        knecht@ipncls.in2p3.fr
S.Krewald       kph101@aix.sp.kfa-juelich.de
M.Lutz          lutz@ect.unitn.it
J.Matias        matias@padova.infn.it
F.Meier         holzwarth@hrz.uni-siegen.d400.de
U.Meissner      meissner@pythia.itkp.uni-bonn.de
B.Moussallam    moussallam@ipncls.in2p3.fr
G.Mueller       mueller@itkp.uni-bonn.de
E.Pallante      pallante@nbivms.nbi.dk
M.Pennington    m.r.pennington@durham.ac.uk
A.Perez         perez@physunc.phy.uc.edu
T.Pham          pham@orphee.polytechnique.fr
J.Portoles      portoles@axpna1.na.infn.it
J.Prades        prades@goya.ific.uv.es
E.de Rafael     rafel@frcpn11.in2p3.fr
M.Sainio        sainio@phcu.helsinki.fi
J.Schechter     schechter@suhep.phy.syr.edu
S.Scherer       scherer@kph.uni-mainz.de
N.Scoccola      SCOCCOLA@MILANO.infn.it
A.Smilga        smilga@vitep5.itep.ru
R.Springer      rps@phy.duke.edu
J.Steele        jim@liszt.physics.sunysb.edu
J.Stern         stern@ipncls.in2p3.fr
D.Toublan       dtoublan@butp.unibe.ch
R.Urech         ru@ttpux2.physik.uni-karlsruhe.de
G.Valencia      valencia@iastate.edu
U.Van Kolck     vankolck@phys.washington.edu
T.Waas          thomas.waas@physik.tu-muenchen.de
T.Watabe        watabe@hadron.tp2.ruhr-uni-bochum.de
H.Weigel        weigel@sunelc1.tphys.physik.uni-tuebingen.de
C.Wiesendanger  wie@stp.dias.ie
A.Wirzba        wirzba@crunch.ikp.physik.th-darmstadt.de
\end{verbatim}

\newpage
\section{The program as it finally ended}
\newcommand{\week}[1]{{\large\bf #1}\\}
\newcommand{\tag}[1]{{\bf #1}\\}
\newcommand{\topic}[1]{{\em #1}\\}
\newcommand{\niks}[1]{}
\begin{tabbing}
14.00 \= a very long name and a\= title of the talk\kill\\
\week{week 1}
\tag{Monday 29/4}
\> morning \> empty for arrival purposes\\

14.30 \>J.~Bijnens \> Administrative and Other Arrangements\\
15.00\> U.~Mei{\ss}ner \> Introductory remarks about CHPT\\
\tag{Tuesday 30/4}
9.15\> Paul B\"uttiker   \> The Chiral coupling constants $\bar{l}_1$ and
$\bar{l}_2$ from $\pi\pi$\\
10.15\>Joseph Schechter \> Simple description of $\pi\pi$ scattering to 1 GeV\\
11.00 \> Coffee Break\\
11.30\>Bachir Moussallam \> Sum rules in $\pi\pi$ scattering\\
12.00\>Lunch\\
15.00\> Veronique Bernard\>$\pi N\to\pi\pi N$ in CHPT\\
15.45\>Stefan Scherer \> Extension of the CHPT Meson Lagrangian to
Order $p^6$\\
16.30\>Coffee Break\\
17.00\>Eugene Golowich \> Two-loop analysis of Vector-Current and
Axialvector-current\\
      \>          \> Propagator, a progress report.\\
17.45\\

\tag{Wednesday 1/5}
10.00\>Ubirajara van Kolck \> Isospin as an Accidental Symmetry\\
10.45\>Coffee Break\\
11.15\>Christoph Hanhart \> $\pi$-threshold production in pp-collisions\\
11.45 \>Lunch\\

15.00\>
Frank Meier  \> Quantum Corrections to Baryon Properties in Chiral \\
       \>          \> Soliton Models\\
15.30 \>
Andreas Wirzba\> S-wave pion propagation in dense isosymmetric nuclear matter\\
16.15 \>Coffee Break\\
16.45 \>
Jan Stern \> Experimental signature of quark condensates\\
17.30\\

\tag{Thursday 2/5}
9.15 \> J\"urg Gasser    \> Pion polarizabilities to two loops\\
10.15 \> Coffee Break \\
11.00\>Tri Nang Pham \> Chiral Lagrangian with vector and axial vector mesons
for\\
            \>  \> $\pi^+$-$\pi^0$ mass difference\\
11.30\>Antonio Perez\> Electromagnetic mass differences of pions and kaons\\
12.00\>Lunch\\
15.00\>Matthias Lutz \>Chiral symmetry and many-nucleon systems\\
15.45 \>Guido M\"uller   \> Renormalization of the Pion-Nucleon Lagrangian to
order $p^4$\\
16.15 \> Coffee Break\\
16.45\>
Norbert Kaiser \>  Neutral pion photo-- and electroproduction\\
17.30\\

\tag{Friday 3/5}
9.15\>Dieter Dreschel\> Pion Photoproduction of the Nucleon -results
from \\ \> \>  Dispersion Theory\\
10.00\>Bugra Borasoy \> Baryon Masses to Second Order in the Quark Masses\\
10.45\>Coffee Break\\
11.15\>James V. Steele \> Master Approach in the Nucleon Sector\\
12.00\>Lunch\\
15.00\>
A. Smilga \> Scalar susceptibility in QCD and in multiflavor Schwinger model\\
15.45\>Jan Stern\>Quark condensate and density of states\\
16.15\\

\week{week 2}
\tag{Monday 6/5}
9.45\>Gilberto Colangelo \> Elastic $\pi\pi$ scattering to Two Loops\\
10.30\> Coffee Break \\
11.00\>Marc Knecht    \> The $\pi\pi$ scattering amplitude to two loops\\
11.45\>Dominique Toublan \> Low Energy Sum Rules For Pion-Pion Scattering\\
\> \>
and Threshold Parameters\\
12.15\>Lunch\\
15.00\>
Herbert Weigel \> Heavy Quark Solitons \\
15.45\>Coffee Break\\
16.15\> Eduardo de Rafael\>Low Energy QCD in the large $N_c$ limit\\
17.00\\

\tag{Tuesday 7/5}
9.15\>Michael Pennington \> Dispersive analysis of $\chi$PT predictions\\
10.00\>  Christian \> Final State Interactions and Khuri-Treiman Equations\\
         \> Wiesendanger              \> in $\eta\rightarrow 3\pi$\\
10.45\>Coffee Break\\
11.15\>
Thomas Hemmert \> $\Delta(1232)$ in Chiral Perturbation Theory\\
12.00\>Lunch\\
15.00\>
Joachim Kambor \>Resonance Saturation in the Baryon Sector\\
15.45\>Coffee Break\\
16.15 \>Silas Beane\> Novel algebraic consequences of chiral symmetry\\
17.00\\
\tag{Wednesday 8/5}
9.15\>
Elisabetta Pallante \> Hadronic Contributions to the muon g-2: an updated
analysis\\
10.00\>Joaquim Prades \> Some Hadronic Matrix Elements in ENJL :\\
\>\>$B_K$, Dashen's Theorem, $\gamma\gamma \rightarrow \pi\pi$ \\
10.45\>Coffee Break\\
11.15\>Res Urech      \> On the corrections to Dashen's theorem\\
11.45\>Lunch\\

\tag{Thursday 9/5}
9.15\>
Giancarlo D'Ambrosio \> Topics in Radiative Non-leptonic Kaon Decays\\

10.00\>Gino Isidori   \> Radiative Four-Meson Amplitudes\\
10.45\>Coffee Break\\
11.15\>Gerhard Ecker  \> Aspects of renormalization in CHPT\\
12.00\>Lunch\\
15.00\>Roxanne P. Springer \> Chiral Symmetry and Hypernuclei\\
15.45\>Norberto Scoccola\>Hyperon Electromagnetic Properties in a Soliton
Model\\
16.15\> Coffee Break\\
16.45\>Teruaki Watabe  \> Strange Contents in Nucleon; Difficulty and
Approach\\
17.30\\
\tag{Friday 10/5}
9.15\>Joaquim Matias \> CHPT description of the MSM: One and two loop order\\
10.00\>Stephan Duerr  \> The covariant derivative expansion\\
10.30\>Coffee Break\\
11.00\>Thomas Waas\> Kaon nucleon interaction and the\\
\>\>$\Lambda(1405)$ in dense matter\\
11.45\>\\
14.00\> Johan Bijnens \> $\gamma\gamma\to\pi\pi\pi$ and some comments
on $U(1)_A$.\\
14.30\>Ulf Mei{\ss}ner\> The organizers have the final word as usual.
\end{tabbing}
\newpage 
\begin{center}
{\large\bf Remarks on CHPT and EFTs}\\[0.5cm]
{\bf Ulf--G. Mei{\ss}ner}\\
Universit\"at Bonn, Institut f\"ur Theoretische Kernphysik\\
D-53115 Bonn, Germany
\end{center}

In this introductory talk, I discuss certain aspects of chiral
perturbation  theory (CHPT), which is the effective field theory (EFT)
of the standard model. In EFTs, based on the power counting first
introduced by Weinberg [1], one considers tree and loop graphs
of the light dofs (the heavy ones being integrated out) and the
effective Lagrangian is organized according to the {\it chiral}
{\it dimension} (or number of derivatives). Often it is argued that
it makes no sense to consider
loops in such an approach since the loop momenta can not be considered
small as it is the case for the external momenta. Clearly, one could
invent a scheme in
which one would cut all the pion momenta to be less than the typical
scale of the heavy degrees of freedom. Alternatively, using
dimensional regularization, one chooses the associated scale to be of
the order of the heavy mass scale. This effectively suppresses the
high momentum components in the loops. However, this high--energy
information is not lost, it is encoded in the values of the associated
low energy constants (LECs) appearing to the order one is working.

A second remark concerns the use of dispersion relations to not only
extend the range of applicability of the chiral predictions but also
to sharpen these at the very low energy end. Having calculated
e.g. the imaginary part of the scalar pion form factor and the
elastic $\pi \pi$ scattering amplitude to one loop [2] allows one
to write a dispersive representation to two loop accuracy for the
scalar form factor $\Gamma_\pi$ with a
number of subtractions to guarantee convergence [3]. These subtraction
constants play a role similar to the LECs in the corresponding
``real'' two loop calculation. While the analytic structure of the two
approaches is identical, the latter one contains more
information. First, the LECs can be related to other processes and
second, only in certain circumstances the enhancement of certain LECs
due to IR logs can be unraveled in the dispersive approach, e.g. the
finiteness of the $\Gamma_\pi$ in the chiral limit reveals the $\ln
M_\pi^2$ dependence of $\bar{d}_2$. Terms of the type $M_\pi^2 \ln
M_\pi^2$ can not be found by such means. If one only considers a
certain observable or process, like $\Gamma_\pi$ or
$\pi\pi \to \pi \pi$, a next to
next to leading order calculation is certainly much easier done in the
dispersive approach. Also,  the equivalent of a one loop calculation can be
done without ever performing a loop integral. The prize one pays is
the loss of information described above.

{\bf References}\\
{} [1] S. Weinberg, Physica 96A (1979) 327\\
{} [2] J. Gasser and H. Leutwyler, Ann. Phys. (NY) 158 (1984) 142\\
{} [3] J. Gasser and Ulf--G. Mei{\ss}ner, Nucl. Phys. B357 (1991) 90.
\newpage 
%
%
%
\newcommand{\al}{\mbox{$\alpha$}}
\newcommand{\B}{\mbox{$\beta$}}
\newcommand{\G}{\mbox{$\gamma$}}
\newcommand{\Fpi}{\mbox{$F_\pi$}}
\newcommand{\lb}[1]{\mbox{$\bar{l}_#1$}}
\newcommand{\pipi}{\mbox{$\pi\pi\mbox{ }$}}
\begin{center}
{\large\bf Chiral Coupling Constants from \pipi Phase Shifts}
\vskip 0.5cm
{B.~Ananthanarayan and \bf P.~B\"uttiker} \\
Institut f\"ur Theoretische Physik, Universit\"at Bern, CH--3012 Bern,
Switzerland

\end{center}
ChPT [1] provides the low energy effective theory of the
standard model describing interactions involving hadronic degrees of
freedom. It is a nonrenormalizable theory; additional coupling
constants have to be introduced at each order of the derivative or
momentum expansion. At leading order $O(p^2)$ there are two such constants,
the pion decay constant $F_\pi$ and the pion mass $m_\pi$. At next to leading
order $O(p^4)$ there are ten more constants. Four of them \lb{1}, \lb{2},
\lb{3} and \lb{4} enter the \pipi scattering amplitude. As a result, the
threshold parameters can be expressed in terms of these as well.
In the past the coupling constants \lb{1} and \lb{2} have been fixed from
experimental values for the D-wave scattering lengths [2] or from an analysis
of $K_{l4}$ decays [3].
On the other hand \pipi scattering has been studied in great detail in
axiomatic field theory [4]. Fixed-t dispersion relations have been established
in the axiomatic framework and properties of crossing and analyticity have
been exploited to establish the Roy equations, a system of integral equations
for the partial wave amplitudes [5,6].
Here we report on a direct determination
of the coupling constants from the existing phase shift data [7,8] by
performing a Roy equation fit to it when $a^0_0$ is restricted to the range
predicted by ChPT. Using certain properties of the chiral amplitude [9], we
write down a dispersion representation with a certain number of subtractions
consistent with $O(p^4)$ accuracy, where the subtraction constants are
expressed in terms of the
chiral coupling constants. The fixed-t dispersion relations of axiomatic field
theory are also rewritten in a form whereby a direct comparison with the
chiral dispersive representation can be made while the subtraction constants
are now computed in terms of physical partial waves, produced by the Roy
equation fit. As an example we cite $\lb{1} = -1.7\pm 0.15$ and
$\lb{2}\approx 5.0$ for the one-loop coupling constants. Our method is
powerful enough to be extended in a straight forward manner to determine
two-loop coupling constants.

{\bf References}\\
{}[1] J.~Gasser and H.~Leutwyler, Ann. Phys. (N.Y.) {\bf 158}, 142 (1984)\\
{}[2] M.~M.~Nagels {\it et al.}, Nucl. Phys. {\bf B 147}, 189 (1979)\\
{}[3] C.~Riggenbach, J.F.~Donoghue, J.~Gasser and B.R.~Holstein,
      Phys.Rev. {\b43},127 (1991); J.~Bijnens, G.~Colangelo and
      J.~Gasser, Nucl.~Phys. {\bf B427}, 427 (1994)\\
{}[4] A.~Martin, ``Scattering Theory: Unitarity, Analyticity and Crossing",\\
\phantom{{}[4]}      Springer-Verlag, Berlin, Heidelberg, New York, 1969\\
{}[5] S.M.~Roy, Phys. Lett. {\bf 36B}, 353 (1971)\\
{}[6] J.--L.~Basdevant et al., Nucl. Phys. {\bf B72}, 413 (1974)\\
{}[7] W.~Ochs, Thesis, Ludwig-Maximilians-Universit\"at, M\"unchen, 1973\\
{}[8] L.~Rosselet, et al., Phys. Rev. {\bf D 15}, 574 (1977)\\
{}[9] J.~Stern, H.~Sazdjian and N.~H.~Fuchs, Phys.~Rev. {\bf D 47},
      3814 (1993)
\newpage 
\begin{center}
{\large {\bf Simple Description of $\pi\pi$ Scattering to One GeV}}\\[0.5cm]
 M. Harada, F. Sannino and {\bf J. Schechter}\\
{ Department of Physics}\\
{ Syracuse University}\\
{ Syracuse, NY 13244-1130, USA}\\
\end{center}
    In this work, described in detail in [1], we slightly relax the extremely
accurate description of the threshold region obtained in chiral perturbation
theory in order to describe $\pi\pi$ scattering all the way up to the 1 GeV
region.

    The present model can be viewed as an attempt to approximate the leading
``Born'' term of the $1/N_c$ expansion of the $\pi\pi$ amplitude in QCD. It
is known that such an amplitude contains contact terms and an infinite number
of resonance exchanges. We truncate the resonances to those in the energy
region up to about 1.4 GeV. We get the amplitude from a chiral Lagrangian so
that crossing symmetry is automatically satisfied. Since the leading $1/N_c$
amplitude contains singularities (zero width resonances) and is otherwise
purely real we a) restrict attention to predicting the real part of the
amplitude b)regularize the amplitude at the pole positions in such a way that
``local unitarity''(near the resonance poles) is maintained.

     It is found that the resulting amplitude satisfies the unitarity bounds
in addition to crossing symmetry for the $I=l=0$ channel (the difficult one)
up to 1.2 GeV. The components are 1) the ``current algebra'' contact term,
2) the $\rho$ exchange diagrams, 3) a broad scalar resonance at about 560 MeV,
and 4) the $f_0(980)$ with its associated Ramsauer-Townsend effect. The
contributions of the ``next group'' of resonances (comprising the
$f_2(1275)$,the
$f_0(1300)$ and the $\rho(1450)$ tend to cancel each other and thus do not
disturb the nice picture in this energy range.
    A similar mechanism is observed in the off diagonal process $\pi\pi
\rightarrow K \bar{K}$.

{\bf References}\\
   {}[1] M. Harada, F. Sannino and J. Schechter, hep-ph/9511335. See also
F. Sannino and J. Schechter, Phys Rev D{\bf{52}},96(1995).
\newpage 
\begin{center}
{\large\bf Determination of Two-loop $\pi\pi$ Scattering Amplitude
Parameters}\\[0.5 truecm]
{\bf Bachir Moussallam}\\
\setcounter{footnote}{0}
{\sl Division de Physique Th\'eorique\footnote{Laboratoire de Recherche des
Universit\'es Paris XI et Paris VI associ\'e au CNRS}, Institut de Physique
Nucl\'eaire}\\
{\sl F-91406 Orsay C\'edex, France}\\
\end{center}

The expansion of the $\pi\pi$ amplitude to two-loop chiral order has
recently been worked out [1][2], it is expressed in terms of elementary
functions of the Mandelstam variables and of six parameters. These involve
combinations of $O(p^4)$ as well as $O(p^6)$ low energy constants and chiral
logarithms [2]. We have shown that four of these parameters obey sum rules
which allow an accurate determination based on existing $\pi\pi$ scattering
data at medium energies ($\sqrt{s}$ between 0.5 to 2 GeV)[3]. These sum rules
are obtained by matching chiral perturbation theory with dispersion
relations, a technique which was used in a variety of applications
in recent years. A specific feature of the elastic $\pi\pi$ amplitude,
is the
invariance under crossing (modulo a crossing matrix). This invariance gives
rise to the Roy dispersive representation [4]. We have rederived and improved
this representation taking into account the notion of chiral counting,
dropping all contributions of chiral order higher or equal to eight. As a
consequence, all the necessary and sufficient conditions for crossing symmetry
to hold can be explicitated in a simple way
(which was not the case in the original formulation): in addition to
determining
the subtraction functions up to two constants, the so called driving terms
are also determined to be polynomials and, finally, a relation is found
between three integrals over high-energy data. Equating the scattering
function $A(s,t,u)$ from a) the chiral expansion and b) the Roy dispersive
representation  gives the four sum rules. We have finally shown
that it is by no means necessary to solve numerically the Roy equations in
order to exploit the sum rules: in the low energy region, where sufficiently
precise scattering data is not available, it is as efficient to use the chiral
expansion of the amplitude: taking two parameters as input ( equivalently, one
could use $a_0^0$ and $a_0^2$) the four remaining ones are determined in a self
consistent way by the sum rules. One of the two parameters which are left free
has a particularly interesting theoretical significance related to the way
in which chiral symmetry is spontaneously
broken in QCD: the interested reader should consult
the contribution of Marc Knecht in these proceedings.\\

{\bf References}\\
{}[1] M. Knecht, B. Moussallam, J. Stern and N. H. Fuchs, Nucl. Phys. B457
(1995) 513.\\
{}[2] J. Bijnens, G. Colangelo, G. Ecker, J. Gasser, M. Sainio,
Phys. Lett. B374 (1996) 210.\\
{}[3] M. Knecht, B. Moussallam, J. Stern and N. H. Fuchs, hep-ph 9512404,
to appear in Nucl. Phys. B. \\
{}[4] S.M. Roy, Phys. Lett. 36B (1971) 353, J.L. Basdevant, J.C. Le Guillou
and H. Navelet, Nuovo Cimento 7A (1972) 363.
\newpage 
\begin{center}
{\large\bf The reaction $\pi N \to \pi\pi N$ in HBCHPT}\\[0.5cm]
{\bf V\'eronique Bernard}, Norbert Kaiser, Ulf--G. Mei{\ss}ner\\
Universit\'e Louis Pasteur, Physique Th\'eorique\\
F-67037 Strasbourg Cedex 2, France
\end{center}

In the framework of heavy baryon chiral perturbation theory (HBCHPT),
we give the chiral expansion for the $\pi N \to \pi \pi N$ threshold amplitudes
$D_1$ and $D_2$ to linear and to  quadratic order in the pion mass.
To linear order in the pion mass, we derive
low--energy theorems (LETs) for the two threshold amplitudes $D_1$ and $D_2$
which are free of unknown low--energy constants. The numerical
predictions of these LETs  work well for the reaction
$\pi^+ p \to \pi^+ \pi^+ n$ but show some significant deviations
for $\pi^- p \to \pi^0 \pi^0 n$ as naively expected [1]. To second
order in the pion mass, the theoretical results  agree within one
standard deviation with the empirical values [2]. We notice that the
effect of pion rescattering is efficiently masked by pion--nucleon
rescattering and resonance excitation, in particular due to the
$N^\star (1440)$. We find a novel $N^\star \to N (\pi \pi)_S$
coupling which has not been accounted for in previous phenomenological
analysis. We also derive a relation between the two threshold amplitudes of
the reaction $\pi N \to \pi \pi N$ and the $\pi \pi$ S--wave scattering
lengths, $a_0^0$ and $a_0^2$, respectively, to order ${\cal
  O}(M_\pi^2)$ [2].
We show that the uncertainties mostly related to resonance excitation
 make an accurate determination of the $\pi \pi$
scattering length $a_0^0$ from the $\pi \pi N$ threshold amplitudes at present
very difficult. From the existing data, we deduce $a_0^0 = 0.21 \pm
0.07$ where the error does not include (presumably large)
contributions at ${\cal O}(M_\pi^3)$.
 The situation is different in the $\pi \pi$ isospin two
final state. Here, the chiral series converges and one finds $a_0^2 = -0.031
\pm 0.007$ somewhat smaller than the two--loop chiral perturbation
theory prediction.
These results could be used to determine $l_3$ which is the parameter directly
related to the size of the condensate. However, at the present state of the art
$l_3$ will be given with a rather large error bar.
We also point out that previous analysis of the
same data using the Olsson--Turner model can not be trusted [3].

{\bf References}\\
{} [1] V. Bernard, N. Kaiser and U. Mei\ss ner, Phys. Lett. B332
(1994) 415; (E) B338 (1994) 520.\\
{} [2] V. Bernard, N. Kaiser and U. Mei{\ss}ner, Nucl. Phys. B457 (1995) 147.\\
{} [3] M.G. Olsson, U. Mei\ss ner, N. Kaiser and V. Bernard, $\pi N$
Newsletter 10 (1995) 201
\newpage 
\begin{center}
{\large\bf Extension of the chiral perturbation theory meson Lagrangian
to order $p^6$}\\[0.5cm]
 Harold W.\ Fearing (1) and {\bf Stefan Scherer (2)}\\
(1) TRIUMF, Vancouver, Canada\\
(2) Institut f\"ur Kernphysik, Mainz, Germany
\end{center}
   We discuss the most general chirally invariant Lagrangian
${\cal L}_6$ for the meson sector at order $p^6$ within the framework
of standard SU(3) chiral perturbation theory [1].
   The result [2] provides an extension of the well-known Gasser-Leutwyler
Lagrangian ${\cal L}_4$ to one higher order, including as well all the
odd-intrinsic-parity terms in the Lagrangian.
   We have developed a systematic strategy so as to get all the independent
terms and eliminate the redundant ones in an efficient way.
   For that purpose we have introduced a twofold hierarchy in terms
of a) the number of covariant derivatives and b) the number of traces
contained in an expression.
   This procedure allows to eliminate terms in favor of ones lower
in the hierarchy without actually working out the explicit and often extremely
complicated relations connecting the corresponding terms.
   We explain how field transformations can be used to identify redundant
terms which are proportional to the lowest-order equation of motion
[3].
   The claim to have obtained the most general Lagrangian relies on
this systematic construction and on the elimination of the redundant
quantities using using relations of which we are aware,
rather than on a general formal proof of either completeness or
independence.

   The end result involves more than hundred terms which, under certain
assumptions, fall into two distinct classes of interaction terms,
according to whether they are even or odd in the number of Goldstone
bosons.
   We have separated the final set of terms into groupings of expressions
contributing to increasingly more complicated processes, so that one
does not have to deal with the full result when calculating $p^6$
contributions to simple processes.

{\bf References}\\
{}[1] J. Gasser and H. Leutwyler, Nucl. Phys. B250 (1985) 465.\\
{}[2] H. W. Fearing and S. Scherer, Phys. Rev. D53 (1996) 315.\\
{}[3] S. Scherer and H. W. Fearing, Phys. Rev. D52 (1995) 6445.
\newpage 
\begin{center}
{\large\bf Two-loop Analysis of Vector-current and Axialvector-current
Propagators in Chiral Perturbation Theory:  A Progress Report.}\\[0.5cm]
{\bf Eugene Golowich}\\
University of Massachusetts\\
Amherst MA 01003 USA
\end{center}

In this talk, I describe a calculational program by Joachim Kambor
and myself for determining the low energy behaviour of
propagators $\Delta_{{\rm V,A},33}^{\mu\nu}(q^2)$ and
$\Delta_{{\rm V,A},88}^{\mu\nu}(q^2)$ at two-loop order in ChPT.

The vector-current part of the project was recently brought to a
satisfactory conclusion [1,2].  We determined the propagators with
straightforward Feynman diagram methods by making appropriate use
of external vector sources. At ${\cal O}(q^4)$ there were three
diagrams and at ${\cal O}(q^6)$ there were ten.  To absorb divergences
and scale dependence at two-loop level required construction of
appropriate counterterms from the ${\cal O}(q^6)$ chiral lagrangian.
Of the four such ${\cal O}(q^6)$ counterterms found, only three
are independent in the vector-current sector.  The final results,
in finite, covariant, and scale-independent form, yielded a
successful fit to data of the two-loop isospin vector spectral
function for $E \le 400$~MeV. [1]

In [2], we used `inverse-moment' sum rules derived in [1]
to obtain phenomenological evaluations of two of the three new
${\cal O}(q^6)$ counterterms.  Our analysis also yielded insights
on the important but difficult issue regarding contributions of
higher orders in the ChPT expansion to the sum rules.

Work continues [3], now on the axialvector propagators.  Although
the vector and axialvector calculations are similar in overall
structure, there are several differences of detail, {\it e.g.} two-loop
renormalizations of masses and decay constants appear in the
axialvector systems via the meson pole.  To date, we have determined
all one-particle irreducible two-loop diagrams, including the
formidable `sunset' diagram.  We have also obtained the list
of ${\cal O}(q^6)$ counterterms which contribute to the
axialvector sector.

{\bf References}\\
{}[1] E. Golowich and J. Kambor, Nucl. Phys. {\bf B447}, (1995) 373.\\
{}[2] E. Golowich and J. Kambor, Phys. Rev. {\bf D53}, (1996) 2651.\\
{}[3] E. Golowich and J. Kambor, work in progress.
\newpage 
\begin{center}
{\large\bf Isospin as an Accidental Symmetry}\\[0.5cm]
{\bf U. van Kolck}\\
Department of Physics, University of Washington\\
Seattle, WA 98195-1560, USA
\end{center}
Isospin is shown [1] to be an accidental symmetry, in the sense of being
a symmetry that is present in the effective low-energy theory (the general
chiral lagrangian involving pions, nucleons, and delta isobars)
{\it in lowest order} but not in the underlying theory ($QC$+$ED$).

I start by constructing the operators involving the
low-energy degrees of freedom that break chiral symmetry in the same way as
quark mass and charge difference terms in $QC$+$ED$; such operators
appear in the chiral lagrangian with coefficients proportional
to powers of the up-down mass difference and the fine structure constant.
I use naive dimensional power counting
to show that there are no isospin violating operators in
lowest order: in most processes isospin violation will therefore
be down compared to isospin conserving terms not only by a ratio of the quark
mass difference to the sum, but also by additional powers of the ratio between
the low energy
of interest ($Q \sim$ pion mass) and the QCD scale ($M \sim$ rho mass).
The operators are next used to study simple processes in leading orders.
It is easy to see that isospin violation in pion-pion
scattering should come predominantly from explicit photon
loops.
Pion-nucleon scattering, on the other hand, could
in principle show ``large'' ($i.e.$, not suppressed by extra powers
of $Q/M$) isospin violation related to quark mass effects in the
$t$-channel isoscalar amplitude, but this is hard to measure.
Turning to nuclear systems, I find that
$i)$ the leading breaking of isospin
comes from the (predominantly electromagnetic) pion mass difference in
the one-pion-exchange two-nucleon potential, an effect that still preserves
charge symmetry;
$ii)$ charge symmetry breaking is $O(Q/M)$ smaller, and arises from an
isospin-violating pion-nucleon coupling, and two short-range
interactions, all mainly quark mass effects. (From the viewpoint
of meson-exchange models, they might come respectively from $\pi$--$\eta$
mixing, and $\rho$--$\omega$ and $a_1$--$f_1$ mixings [2].)
Finally, the use of such a chiral lagrangian is illustrated by a computation
of the pion-range, isospin-violating two-nucleon potential to
third order in the chiral
expansion [3]. This includes, besides the tree graphs with isospin dependent
pion masses and couplings mentioned above, also one-loop diagrams
with pion [2] and photon dressings at vertices and propagators,
and with simultaneous pion and photon exchange.

{\bf References}\\
{}[1] U. van Kolck, Ph.D. Dissertation, Univ. of Texas (1993);
                    Univ. of Washington preprint DOE/ER/40427-13-N94
                    (in preparation).\\
{}[2] U. van Kolck, J.L. Friar, and T. Goldman, Phys. Lett. B 371 (1996) 169.\\
{}[3] J.L. Friar, T. Goldman, and U. van Kolck, in preparation.
\newpage 
\begin{center}
{\large\bf Threshold pion production in nucleon-nucleon collisions}\\[0.5cm]
{\bf C. Hanhart}, J. Haidenbauer and J. Speth\\
Institut f\"ur Kernphysik, Forschungszentrum J\"ulich\\
J\"ulich, Germany\\
\end{center}

We performed a momentum-space calculation of
the reaction $NN \to NN\pi$ near threshold, extending our earlier study [1].
The following pion production mechanisms
are considered: (i) Direct emission of the pion form one of the nucleons.
(ii) (s-wave) rescattering where the produced pion first scatters off the other
nucleon
before its emission. (iii) Contributions from meson-exchange currents due to
the exchange
of heavy mesons ($\sigma, \ \omega$) connected to intermediate
$N\bar N$ pairs[2].
The Bonn OBEPT potential [3] is used for the distortions in the initial
and final $NN$ states. For the evaluation of the rescattering contribution a
microscopic
meson--exchange model of $\pi N$ interaction developed recently by the J\"ulich
group [4]
is utilized.
A soft form factor (of monopole type) with a cut--off mass $\Lambda_{\pi NN} =
800$ MeV,
as suggested, e.g., by recent QCD lattice calculations [5], is employed at the
pion production
vertex. In the calculation of the heavy-meson-exchange contributions we us the
same
$\omega$ vertex parameters as in OBEPT. The $\sigma$, however, is an effective
parameterization
of correlated 2-$\pi$exchange (and other processes) and its strength should be
different in the NN interaction and in the present case, where it couples to
$N \bar N$ pairs. Therefore the coupling constant of the $\sigma$ is treated as
free parameter.
We achieve a quantitative description of the measured total cross section
for the reaction $pp \to pp \pi^0$ near threshold. With the same model (and the
same parameter set) we are also able to reproduce th $pp \to d \pi^+$ cross
section
near threshold with similar quality. Note, however, that in the latter reaction
the pion production
via a $\Delta$-isobar excitation could be important [6]. This process is so far
neglected in
our study.
The reaction $NN \to NN\pi$ close to threshold promises to
give a deeper insight into the off--shell properties of the $NN$ interaction as
well as into the short range correlations of the NN force.
It might also be sensitive to additional constrains from chiral symmetry
as suggested by recent investigations  using chiral perturbation theory[7].

{\bf References}\\
{}[1] C. Hanhart, J. Haidenbauer, A. Reuber, C. Sch\"utz
 and J. Speth, Phys. Lett. {\bf B358}(1995)21-26.\\
{}[2] T.-S. Lee and D. Riska, Phys. Rev. Lett. {\bf 70},  2237  (1993).\\
{}[3] R.Machleit, K.Holinde, and Ch.Elster, Phys. Rep. {\bf 149}, 1 (1987) \
.\\
{}[4] C. Sch{\"u}tz, J.W. Durso, K. Holinde, and J.Speth, Phys. Rev. C
  {\bf 49},  2671  (1994).\\
{}[5] K.F.Liu, S.J.Dong, T.Draper, and W.Wilcox, Phys. Rev. Lett. {\bf 74},
2171 \ . \\
{}[6] J.A.Niskanen, Phys. Rev. C, {\bf 53}, 526 (1996) \ . \\
{}[7] T.Y.Park et al., Phys. Rev. C, {\bf 53}, 1519 (1996) and
T.D.Cohen et al., nucl-th/9512036 .
\newpage 
\begin{center}
{\large\bf S-wave pion propagation in dense isosymmetric nuclear
matter}\\[0.5cm]
{\bf Andreas Wirzba}\\
Institut f\"ur Theoretische Kernphysik, Technische Hochschule Darmstadt\\
Schlo{\ss}gartenstr.~9, D-64289 Darmstadt, Germany
\end{center}
The starting point of the talk is the isoscalar $S$-wave interaction
between pions and nucleons.  The corresponding ${\cal
O}(Q^2)$-heavy-baryon lagrangian of ref.~[1] (in the mean-field
approximation for the nucleons) is then applied to finite baryon
densities $\rho$. Note that nuclear correlations are thus neglected.
Within the generating functional formalism of ref.~[2] the effective
mass of the pion in isosymmetric homogeneous nuclear matter is derived
[3,4] and shown to be independent of the various off-mass-shell
extension schemes (as e.g.\ PCAC) [3].

With the help of the corresponding generating functional the
density-dependent quark condensate $\langle \bar u u\mbox{+}\bar d
d\rangle_\rho$ as well as the in-medium axialvector-axialvector and
axialvector-pseudoscalar correlators are derived and the time-like
pion decay constant $F_\pi^t(\rho)$ (which in the matter background is
bigger than its space-like counter part $F_\pi^s(\rho)$) and the
pseudoscalar coupling constant $G^\ast_\pi(\rho)$ are deduced [3,4].
These quantities combine to satisfy the in-medium extension of the
Gell-Mann-Oakes-Renner relation, $ F_\pi^t(\rho)^2 m_\pi^\ast(\rho)^2
= -\hat m_q \langle \bar u u +\bar d d \rangle_\rho +{\cal
O}(Q^3,\rho^2) $, and the density-dependent PCAC relation, $
F_\pi^t(\rho) m_\pi^\ast(\rho)^2 = \hat m_q G_\pi^\ast(\rho) + {\cal
O}(Q^3,\rho^2) $, where $\hat m_q$ is the SU(2)-averaged current quark
mass.  Furthermore the results are compatible with Migdal's approach
to finite Fermi systems on the composite hadron level provided the
Migdal propagator has been identified correctly [5].

Finally, the mean-field lagrangian sets constraints for the in-medium
extension of chiral perturbation theory [4].  As the matter background
selects a special Lorentz-frame, the in-medium version of ChPTh cannot
satisfy Lorentz invariance, but only the left-over Euclidean
rotational invariance. It should therefore be classified as a
non-relativistic ChPTh of ref.~[6].  Indeed, the dispersion of the
$S$-wave pion-propagation in isosymmetric nuclear matter is to order
${\cal O}(Q^2)$ the same as for the corresponding Goldstone ($\pi$)
bosons of an antiferromagnet, where the spacelike ``$F_\pi^s$'' is
smaller than the time-like ``$F_\pi^t$'' [6] as well.  However, the
fact that the $\pi N$ $S$-wave lagrangian predicts, in the mean-field
approximation, the first corrections to the isoscalar channel to be of
order ${\cal O}(Q^3)$ is incompatible with standard ChPTh (where they
are of order ${\cal O}(Q^4)$ whether in the relativistic or
non-relativistic version). Thus the in-medium (non-relativistic) ChPTh
has to be of the generalized form of ref.~[7]. This is supported by
the possibility that, with increasing baryon density, the in-medium
quark condensate can potentially become so small, that the higher
in-medium quark condensates cannot be neglected any longer as it is
the case in standard ChPTh in the vacuum.

{\bf References}\\
{} [1] V. Bernard, N. Kaiser and U.-G. Mei{\ss}ner,
Phys. Lett. B309 (1993) 421\\
{} [2] J. Gasser and H. Leutwyler, Ann. Phys. (NY) 158 (1984) 142\\
{} [3] V. Thorsson and A. Wirzba, Nucl. Phys. A
589 (1995) 633\\
{} [4] A. Wirzba and V. Thorsson, {\em Hirschegg
'95},  GSI, 1995, {\tt hep-ph/9502314}\\
{} [5] M. Kirchbach and
A. Wirzba, Nucl. Phys. A in print, {\tt nucl-th/9603017}\\
{} [6] H. Leutwyler, Phys. Rev D49 (1994) 3033\\
{} [7] M. Knecht and J. Stern, DAPHNE physics handbook (2nd ed.),
{\tt hep-ph/9411253}

\newpage 
\begin{center}
{\large\bf Experimental Signature of Quark-Antiquark Condensation in the
QCD Vacuum}\\[0.5cm]
{\bf Jan Stern}\\
\setcounter{footnote}{0}
Division de Physique Th\'eorique\footnote{Laboratoire de Recherche des
Universit\'es Paris XI et Paris VI, associ\'e au CNRS}, Institut de Physique
Nucl\'eaire\\
F-91406 Orsay Cedex, France
\end{center}
It is generally believed that in QCD, the spontaneous
breakdown of chiral symmetry is a consequence of a strong quark-antiquark
condensation in the vacuum. The condensate parameter $B_0= - {\bar q}q/F^2$
is expected to be sufficiently large to insure that, for actual values of
running quark masses, the expansion of the square of the Goldstone boson
masses is dominated by the first Gell-Mann--Oakes--Renner term. This
assumption is crucial in the standard formulation [1] of CHPT, but sofar,
it has not been tested experimentally. Moreover, a sound theoretical
alternative [2] in which the condensate $B_0$ would be marginal (typically
10 times smaller than believed) or even vanishing, could naturally arise
in QCD and it is not excluded by any existing data. For these reasons an
experimental probe of quark condensation  becomes of fundamental importance.
The best evidence in favour of or against a strong ${\bar q}q$ condensation
would be provided by new high precision low energy pipi scattering
experiments [3]. ( The actual uncertainty, c.f. $a_0^0= 0.26\pm 0.05$
encompasses
both alternatives of a strong and weak quark condensation.) Additional
signature could emerge from the analysis of deviations from the
Goldberger Treimann relations [4], $\eta\to 3\pi$ decays [5],
$\gamma\gamma\to\pi^0\pi^0$ near
threshold, provided corresponding experimental data become more accurate.
The question of the strength of quark condensation could influence on various
(not yet tested ) predictions: estimates of light quark masses [6],
estimates of $\varepsilon '/\varepsilon$ and some issues in the B-physics,
among others.

{\bf References}\\
{}[1] J. Gasser and H. Leutwyler, Nucl. Phys. B250 (1985) 465.\\
{}[2] M. Knecht and J. Stern in The Second Daphne Physics Handbook, eds.
L.Maiani,G.Pancheri and N.Paver, INFN-LNF publication, May 1995.\\
{}[3] N. H. Fuchs, H. Sazdjian and J. Stern, Phys. Lett. B269 (1991) 183;
Phys. Rev.D47 (1993) 3814.\\
{}[4] N. H. Fuchs, H. Sazdjian and J. Stern Phys. Lett. B238 (1990) 380.\\
{}[5] M. Knecht and J. Novotny in preparation.\\
{}[6] J.Stern, M.Knecht and N.H. Fuchs in Proceedings of the Third Workchop
on the Tau-Charm Factory, Marbella (Spain), June 1993, Eds. J. Kirkby and
R. Kirkby, Editions Frontieres (1993).

\newpage 
\begin{center}
{\large \bf Pion polarizabilities to two loops}\\[0.5cm]
{\bf J. Gasser}\\ Institut f\"ur theoretische Physik,
Universit\"at Bern, \\
Sidlerstrasse 5, CH--3012 Bern
\end{center}

I have discussed in my talk the evaluation of the
 pion
electric $(\bar{\alpha}_\pi)$ and magnetic $(\bar{\beta}_\pi)$
polarizabilities in the framework of chiral perturbation theory.
The polarizabilities are obtained by expanding
the Compton amplitude
 near threshold in powers of the photon momenta. The leading term
in this  {momentum expansion}
 is  proportional to the square of the charge of the
pion, whereas
the coefficients of the next--to--leading--order term are determined
by  $\bar{\alpha}_\pi$ and $\bar{\beta}_\pi$. In the
{chiral expansion}, the polarizabilities receive
their leading
order contribution from terms at order $p^4$ [1-3]. At this order, one has
$\bar{\alpha}_\pi +\bar{\beta}_\pi = 0$ ($D$--wave term).
Therefore, to determine the first nonvanishing contribution to
$\bar{\alpha}_\pi +\bar{\beta}_\pi$, a two--loop
evaluation of the Compton amplitude  is needed.

The result of this calculation in the	neutral channel has been
published some
time ago [4], whereas  the evaluation in the charged channel
(which involves
considerably more diagrams, because there is a tree graph
contribution) has been completed recently
   [5].  The two low--energy constants that enter
the polarizabilities at two--loop order have been determined by
resonance saturation in these references.
Chiral	logarithms
contribute in both channels in a nonnegligible manner.
 B\"urgi has also investigated the importance
of various graphs in the $\overline{{\mbox{MS}}}$ scheme [5],
using the sigmamodel parametrization of the $U$--matrix.
 He finds that
the acnode and the box graphs generate substantial contributions
to the polarizabilities.

{\bf References}\\
{}[1] J.F. Donoghue and B.R. Holstein, Phys. Rev. D40 (1989) 2378.\\
{}[2] B.R. Holstein, Comm. Nucl. Part. Phys. 19 (1990) 221.\\
{}[3] J. Bijnens and F. Cornet, Nucl. Phys. B296 (1988) 557.\\
{}[4] S. Bellucci, J. Gasser and M.E. Sainio, Nucl. Phys. B423
(1994) 80.\\
{}[5] U. B\"urgi, Charged pion polarizabilities to two loops,
Bern University Preprint BUTP--96/01, hep--ph/9602421, to
appear in Phys. Lett. B; Char\-ged pion--pair production and
pion polarizabilities to two loops, Bern University
Preprint BUTP--96/02, hep-ph/9602429.
\newpage 
\begin{center}
{\bf  CHIRAL LAGRANGIAN WITH VECTOR
AND AXIAL \\
VECTOR MESON FOR $\pi^{+}- \pi^{0}$ MASS DIFFERENCE} \\
\bigskip
 {\bf {T. N. Pham} \\}
{\em Centre de Physique Th\'eorique, \\
Centre National de la Recherche Scientifique, UPR A0014, \\
Ecole Polytechnique, 91128 Palaiseau Cedex, France}
\end{center}

\vspace{1cm}

 In this talk, I report on a recent work [1]  on a simple
derivation of the
forward virtual Compton scattering off a soft pion target for use in the
$\pi^{+}-\pi^{0}$ electromagnetic mass difference calculation.
References to previous works can be found in this paper.

By using a nonlinear chiral Lagrangian with vector and axial vector
meson incorporated in a modified gauged chiral model,
it is shown that the simple expression for the forward virtual
Compton scattering
on a soft pion  usually obtained from Current Algebra
 can be derived in a simple manner. This shows
also
that the absence of the double pole behaviour for the Born terms
 is a consequence of chiral symmetry. Though this
result has also been obtained previuosly by Ecker {\em et al.} and also
more recently by Donoghue {\em et al.}
in which the vector and axial vector
meson fields are treated as antisymmetric tensor representation instead
of the usual four-vector field operator, we show that
one can also derive the Current Algebra result
in a simple manner with
the conventional four-vector representation for  vector and axial vector
meson.
 We note also that the
unsubtracted dispersion relation for the $\Delta I = 2$ amplitude can be made
consistent with the soft pion result by including
also the contact term from the vector meson pole term in a modified
Born term. Then it would be more convenient to use the dispersion
relation approach to calculate the  $\pi^{+}-\pi^{0}$ mass difference
since terms of $O(p^{2})$ can also be analysed in a straightforward
manner.

\vspace{3cm}

{\bf References}\\
{[1] T. N. Pham, {\em Phys.Lett.} {\bf 374} (1989) 205 . }
\newpage 
\begin{center}
{\large\bf The Electromagnetic Mass Differences of Pions and
Kaons}\\[0.5cm]
 John F. Donoghue$^{(a)}$ and {\bf Antonio F. P\'{e}rez}$^{(a, b)}$ \\
(a) Department of Physics and Astronomy \\
University of Massachusetts, Amherst, MA ~01003 \\
(b) Department of Physics \\
University of Cincinnati, Cinicinnati, OH
\end{center}

We use the Cottingham method to calculate the pion and
kaon electromagnetic mass
differences with as few model dependent inputs as possible.
The constraints of chiral
symmetry at low energy, QCD at high energy and
experimental data in between are used in
the dispersion relation.  We find excellent agreement with
experiment for the pion mass
difference.  The kaon mass difference exhibits a strong
violation of the lowest order
predictions via Dashen's theorem, in qualitative agreement
with several other recent
calculations.

{\bf References, (Partial List)}\\
{}[1] Aston,~D. {\it et al.}, {\it Nucl. Phys.} {\bf B202,} pg.~21, 1982. \\
{}[2] Baur,~R., Urech~,~R., hep-ph/9508393, Feb. 22 1996. \\
{}[3] Bijnens,~J., {\it Phys. Lett.}, {\bf B306,} pg.~343, 1993. \\
{}[4] Collins,J. C., {\it Nucl. Phys.} {\bf B149,} pg.~90, 1979. \\
{}[5] Cottingham,~W.~N., {\it Annals. Phys.} {\bf 25,} pg.~424, 1963. \\
{}[6] Das et al., {\it Phys. Rev. Lett.} {\bf 18,} pg.~759, 1967. \\
{}[7] Dashen,~R., {\it Phys. Rev.} {\bf 183,} pg.~1245, 1969. \\
{}[8] Daum,~C. {\it et al.}, {\it Nucl. Phys.} {\bf 187,} pg.~1, 1981. \\
{}[9] Donoghue,~J. et al., {\it Phys. Rev.}, {\bf D47,} pg.~2089, 1993. \\
{}[10] Donoghue,~J., and Golowich,~E., {\it Phys. Rev.}, {\bf D49,} pg.~1513,
1994. \\
{}[11] Ecker et al., {\it Nucl. Phys.} {\bf B321,} pg.~311, 1989. \\
{}[12] Ecker et al., {\it Phys. Lett.} {\bf B223,} pg.~425, 1989. \\
{}[13] Gasser,~J. and Leutwyler,~H., {\it Annals of Phys.} {\bf 158,}
pg.~142, 1984. \\
{}[14] Gasser,~J. and Leutwyler,~H., {\it Nucl. Phys.} {\bf B250,}
pg.~465, 1985. \\
{}[15] O'Donnell, P. J., {\it Rev. Mod. Phys.} {\bf 53,} pg.~673, 1981. \\
{}[16] Particle Data Group, {\it Phys. Rev.} {\bf D50,}, 1994. \\
{}[17] Socolow, R. H., {\it Phys. Rev.}
{\bf 137,} pg.~B1221, 1965. \\
{}[18] Weinberg, S., {\it Phys. Rev. Lett.} {\bf 18,} pg.~507, 1967.
\newpage 
\begin{center}
{\large \bf Chiral approach to $NN$-scattering amplitude }\\[0.5cm]
{\bf Matthias Lutz}\\
ECT$^*$ Villa Tambosi\\ Trento, Italy
\end{center}

\noindent
Weinberg [1] and van Kolck [2] introduced a
{\it nonrelativistic} perturbative scheme with consistent chiral
counting rules for the nucleon-nucleon {\it potential}.
In this talk we present a {\it relativistic} chiral expansion scheme for the
nucleon-nucleon scattering {\it amplitude}. It is advantageous to
work with the manifestly Lorenz {\it covariant} chiral Lagrangian where
we achieve the desired nonrelativistic $1/m$ expansion by a proper
regrouping of the interaction terms in
the Lagrangian. The $1/m$ expansion can then be performed at the
level of each individual Feynman diagram. We find that our relativistic
scheme naturally sums the {\it non-polymomial} terms in $1/m$
needed to reconcile proper dispersion relations and threshold behaviour.

Chiral counting rules, predicting the {\it leading } chiral power for the
two-nucleon irreducible Feynman diagrams of the
Bethe-Salpeter {\it kernel}, are in full analogy to Weinberg's counting rules
for the nucleon-nucleon potential. To further derive chiral
power counting rules for the $NN$-scattering {\it amplitude}
we introduce a subtraction scheme (for a given finite cutoff $\Lambda $)
at the level of the Bethe-Salpeter equation such that its solution,
the $NN$-scattering amplitude, is independent of the subtraction scheme with
its characteristic scale $\mu $. A small subtraction scale $\mu $, of the
order of the pion mass, renders the unitary iterations of the properly
subtracted pion exchanges {\it perturbative}
with a well defined chiral counting rule:  each intermediate two-nucleon
state generates a chiral enhancement power -1. The subtraction scale
determines the relative importance of the unitary iterations of
the 2-nucleon vertices as compared to the strength provided by unitary
iterations of pion exchanges. We find that a {\it small} subtraction
scale $\mu $ causes a strong renormalization of the local s-wave
nucleon-nucleon interaction vertices at given physical cutoff $\Lambda $.
In this case the {\it natural} s-wave bare couplings mutate into {\it large}
couplings, which acquire the {\it anomalous} chiral power $-1$. The thus
renormalized local s-wave interaction vertices pick up sufficient
strength to generate naturally  the deuteron bound state and the
pseudo-bound state in the nucleon-nucleon scattering amplitude upon unitary
iterations.

Our scattering amplitude exhibits simple complex pole terms, reflecting
the presence of the pseudo bound state and the deuteron bound state, and
a remainder which comprises the proper cut structure from multiple
pion exchanges. To leading orders the free parameters of our scheme are
in one-to-one correspondence to s-wave scattering lengths and ranges
and p-wave scattering volumes.

\vskip0.2cm
{\bf References}\\
{}[1] S. Weinberg; Nucl. Phys. B363 (1991) 3 \\
{}[2] U. van Kolck; in "{\it Low Energy Effective Theories and QCD}",
    D.-P. Min (ed.), \\
\indent Seoul (1995)
\newpage 
\begin{center}
{\large\bf Renormalisation of the SU(3) chiral meson-baryon lagrangian to order
  q$^3$ }\\ [0.5cm]
{\bf Guido M\"uller} \\
{Institut f\"ur Theoretische Kernphysik}\\
{Nu\ss allee 14-16}\\
{D-53115 Bonn} \\
{e-mail:mueller@pythia.itkp.uni-bonn.de} \\ [0.5cm]
\end{center}
Three-flavor chiral perturbation theory with baryons is a topic of current
interest. The dynamics of kaon-nucleon interactions or
kaon photo(electro)production are  based on the SU(3) extension of
chiral effective lagrangians. We remind that for the isospin-odd pion-nucleon
scattering length the loop correction only can fill the gap between the
Weinberg-Tomozawa prediction and the empirical value. Loops produce in
general ultraviolet divergences, which can be absorbed by introducing
counterterms\,[1]. The knowledge of the full divergence
structure allows to control these calculations.
We perform the complete regularisation of
all Green functions with a single incoming and outgoing baryon to order
q$^3$ in the chiral SU(3) meson-baryon system. The method is based on
the work of
Ecker who performed the complete renormalisation of Green functions of
the pion-nucleon interaction in the heavy baryon formalism\,[2]. This
allows for a consistent chiral power counting. The divergences can be
extracted in a chiral invariant manner by making use of the heat
kernel representation of the propagators in d-dimensional Euclidean
space. The main difference between the two
calculations lies in the fact that the nucleons are in the fundamental
representation of SU(2), while the baryons are in the adjoint
representation of SU(3). This leads to some algebraic
consequences for the construction of the one-loop generating
functional\,[3].

{\bf References}\\
{}[1] J.\,Gasser, M.E.\,Sainio and A.\,Svarc, Nucl. Phys. {\bf B307}
(1988) 779\\
{}[2] G.\,Ecker, Phys. Lett. {\bf B336} (1994) 508\\
{}[3] G.\,M\"uller and U.-G.\,Mei\ss ner, Bonn preprint, in preparation
\newpage 
\begin{center}
{\large\bf Neutral Pion Photoproduction off Protons}\\[0.5cm]
{\bf Norbert Kaiser}\\
Technische Universit\"at M\"unchen, Physik Department T39\\
D-85747 Garching, James-Franck-Stra{\ss}e
\end{center}
I analyse the new threshold data for neutral pion photoproduction off protons
in the threshold region [1,2] within the framework of heavy baryon chiral
perturbation theory at order $q^4$. It is shown that indeed large loop
corrections as predicted by CHPT at order $q^4$ are needed to understand the
small value of the S-wave multipole $E_{0+}$ at threshold [3]. Due to the
rather slow convergence of this quantity in powers of the pion mass it does not
provide anymore a good testing ground of chiral dynamics. However, there are
new and rapidly converging low energy theorems for two combinations of the
P-wave multipoles [3]. The one for $P_1$ holds at the few percent level when
compared to the new TAPS and SAL data [1,2]. The low energy theorem for $P_2$
can be tested soon with polarized photon data taken at MAMI. The few low energy
constants entering the  calculation can be well understood from resonance
saturation [4].

Furthermore, I discuss double $\pi^0$ photoproduction off protons close to
threshold. The chiral loops enter here already at leading nonvanishing order
and considerably enhance the near threshold total cross section.  This feature
remains in a full order $q^4$ calculation of all next-to-leading order
corrections [5].

{\bf References}\\
{} [1] M. Fuchs et al., Phys. Lett. B368 (1996) 20.\\
{} [2] J. Bergstrom et al., Phys. Rev. C53 (1996) R1052.\\
{} [3] V. Bernard, N. Kaiser and U. Mei{\ss}ner, Z. Phys. C70 (1996) 483.\\
{} [4] V. Bernard, N. Kaiser and U. Mei{\ss}ner, "Chiral Symmetry and the
Reaction $\gamma p \to \pi^0 p$", Phys. Lett. B (1996) in print.\\
{} [5] V. Bernard, N. Kaiser and U. Mei{\ss}ner, "Double Neutral Pion
Photoproduction at Threshold", Phys. Lett. B (1996) in print.
\newpage 
\begin{center}
{\large\bf Photoproduction of Pions off the Nucleon -
Results from Dispersion Theory}\\[0.5cm]
{\bf Dieter Drechsel} and Olaf Hanstein\\
Institut f\"ur Kernphysik, Universit\"at Mainz,
D-55099 Mainz, Germany\\
\end{center}
\small
Dispersion relations at constant $t$ [1] have been used to analyze the recent
precision experiments at MAMI (Mainz) and ELSA (Bonn). The partial wave
amplitudes fulfil the phase relations required by the Watson theorem at the
lower energies and some approximate relation at the higher energies. The
contributions of the dispersion integral above $2 GeV$ are replaced by a
fraction of the vector meson exchange representing the Regge behaviour
of the amplitudes expected at very high energies. This unknown high-energy
behaviour and the necessity to add multipoles of the homogeneous equations
to the solutions of the coupled system of multipole amplitudes, leads to the
introduction of 9 free parameters determined by a fit to the data in the
energy region of $160 MeV \le E_{\gamma} \le 450 MeV$.

The threshold region has not been included in our fit because isospin symmetry
breaking effects play an important role in that region due to the different
pion masses. However, the threshold predictions obtained from our
analysis are in very good agreement with the results from chiral perturbation
theory ($ChPT$) [2]. In particular, we find $E_{0+} (n\pi^{+}) = 28.4 \cdot
10^{-3}/m_{\pi}$ and $E_{0+} (p\pi^{-}) = -31.9\cdot10^{-3}/m_{\pi} $.
The latter value is consistent with the angular distribution and the total
cross section of a recent TRIUMF experiment [3] whose analysis  had led to a
threshold amplitude of $E_{0+} (p\pi^{-}) = - 34.7\cdot10^{-3}/m_{\pi}$, in
disagreement with $ChPT$ an
d low energy theorems. It is also interesting that the data at the higher
energies, via dispersion relations, lead to a prediction of $E_{0+} (p\pi^{0})
= - 0.4 \cdot10^{-3}/m_{\pi}$ at $\pi^{+}$- threshold. This is in good
agreement
with both the data [4] and $ChPT$, and a consequence of the importance of loop
corrections to the ''old'' low energy theorem.

In the region of the $\Delta (1232)$ isobar, we have decomposed the $E^{(3/2)}_
{1+}$ and $ M_{1+}^{(3/2)}$ multipoles  into resonance  and background
contributions using the speed-plot technique [5]. In this way we are able
to determine the position of the resonance pole in the complex plane at
$W = M_{R} - i \Gamma_{R}/2$ with $M_{R} = (1211\pm 1) MeV$ and $\Gamma_{R}
= (100\pm 2) MeV$ [6], in excellent agreement with results from
pion-nucleon scattering. The resonant contributions to the two multipoles
 are then determined as the complex residues at the resonance pole,
$R_{E/M}$exp$(i\phi _{E/M})$. While the Watson theorem requires that the
physical amplitudes  $E_{1+}^{(3/2)}$ and $M_{1+}^{(3/2)}$ have the same phase,
the corresponding ratio for the resonant amplitude is a complex number,
$R_{E}$exp$(i\phi_{E})/R_{M}$exp$(i\phi_{M}) = (- .035, - 0.46).$
Recent experiments on angular
distributions and photon asymmetries [7] have found an $E/M$ ratio of
$(- 2.4 \pm 0.2)\%$ at
 $W = 1232 MeV$. Since the experiment is sensitive to the ratio $Re\{E_{1+}
^{\ast} M_{1+}\}/\mid M_{1+}\mid^{2}$, our analysis predicts that the
resonant contribution to the $E/M$ ratio is $-3.5\%$.

In conclusion, dispersion relations at constant $t$ give a
possibility to analyze and to interpret the new precision data on
pion photoproduction. Further improvements are expected to come
from subtracted dispersion relations, by using the threshold results
from $ChPT$ as input, with the consequence of reducing the sensitivity
on the unknown behaviour of the dispersion integrals at the higher energies.\\
{\bf References}\\
{}[1] O. Hanstein, Ph. D. thesis, Mainz (1996).\\
{}[2] V. Bernard et al., Nucl. Phys. B383 (1992) 442, and Z. Phys. C70 (1996)
483.\\
{}[3] Kailin Liu, Ph. D. thesis, University of Kentucky (1994).\\
{}[4] R. Beck et al., Phys. Rev. Lett. 65 (1990) 1841, and M. Fuchs et al.,
Phys. Lett. B368 (1996) 20.\\
{}[5] G. Hoehler, $\pi N$ Newsletter 7 (1992) 94 and 9 (1993) 1.\\
{}[6] O. Hanstein, D. Drechsel and L. Tiator ''The position and the
residues of the delta resonance pole in pion photoproduction'' (to be
published).\\
{}[7] R. Beck, Proc. Int. Conf. ''Baryons '95'', Santa F$\acute{e}$ (1995),
and H.-P. Krahn, Ph. D. thesis, Mainz (1996).
\newpage 
\normalsize
\begin{center}
{\large\bf Baryon Masses and $\sigma$--terms}\\
{\large\bf to second order in the quark masses}\\[0.5cm]
{\bf Bu\=gra Borasoy}
\end{center}
We analyze the octet baryon masses and the pion/kaon--nucleon
$\sigma$--terms in the framework of heavy baryon chiral
perturbation theory, [1]--[4]. We include {\it all} terms up-to-and-including
quadratic order in the light quark masses, $m_q$. The pertinent low--energy
constants are fixed from resonance exchange within the one--loop
approximation. This includes contribution from
loop graphs with intermediate the spin--3/2 decuplet and
the spin--1/2 octet states and from tree graphs including scalar mesons.
We demonstrate that two--loop corrections indeed modify
the leading one--loop results for some of these coefficients. Retaining only
the contributions to the low--energy constants to one--loop order,
the only free parameter is
the baryon mass in the chiral limit, $\stackrel{\circ}{m}$. We find
 $\stackrel{\circ}{m}=
840 \pm 100$~MeV, [5],[6]. While the  corrections of order $m_q^2$ are small
for the nucleon and the $\Lambda$, they are still large for the
$\Sigma$ and the $\Xi$. Therefore a definitive statement about the
convergence of three--flavor baryon chiral perturbation can not yet be
made. The pion--nucleon $\sigma$--term is given parameter--free, we get
$\sigma_{\pi N} (0) = 43 \pm 10$ ~MeV, which is in good agreement with
dispersion-theoretical determinations, together with the
strangeness content of the nucleon, $y = 0.08 \pm
0.12$. We also estimate the kaon--nucleon $\sigma$--terms, the
shifts to the respective Chang--Dashen points and some two--loops
contributions to the nucleon mass.

{\bf References}\\
{}[1]E. Jenkins, Nucl. Phys. {\bf B368} (1992) 190\\
{}[2]V. Bernard, N. Kaiser and Ulf-G. Mei\ss ner, Z. Phys. {\bf C60} (1993)
111\\
{}[3]R.F. Lebed and M.A. Luty, Phys. Lett. {\bf B329} (1994) 479\\
{}[4]M.K. Banerjee and J. Milana, Phys. Rev. {\bf D52} (1995) 6451\\
{}[5]B. Borasoy and Ulf--G. Mei{\ss}ner, Phys. Lett. {\bf B365} (1996)285\\
{}[6]B. Borasoy and Ulf--G. Mei{\ss}ner, Bonn preprint TK-96/14 (1996)
\newpage 
\begin{center}
{\large\bf The Master Formula Approach in the Pion and Nucleon
Sector}\\[0.5cm]
{\bf James V. Steele}\\
SUNY Stony Brook\\
Stony Brook, NY USA
\end{center}
The master formula approach uses an on-shell chiral reduction
scheme in order to make predictions for meson-nucleon dynamics [1].
It relates scattering amplitudes to measurable vacuum correlation
functions hence allowing the inclusion of resonances.  In particular,
it has been applied to $\pi\pi$ scattering where a model independent
verification of the rho dominance in the vector channel was shown [2].

In addition, a semi-classical expansion in $\hbar\sim1/f_\pi^2$ may be
done.  This is equivalent to a momentum expansion in the pion sector.
Two solutions of the master formula result: one reproduces CHPT and
the other allows for an additional relation between the divergences,
reducing the number of parameters to two at one-loop [1,2].

The extension of MF to the nucleon sector was also reported on at this
conference.  The $1/f_\pi$ expansion is no longer equivalent to CHPT
since there are many mass scales involved.  The $\pi$-$N$ sigma term
is found to be given by the Goldberger-Treiman discrepancy to tree
level [3].  Using this one finds naturally that every time $g_A$
appears it is replaced by $g_{\pi NN}$.

A comparison with all form factors and many scattering amplitudes
calculated in the relativistic formulation of chiral perturbation
theory [4,5] shows additional differences.  The scalar form
factor and $\pi N\to \pi\pi N$ scattering amplitude have additional
momentum dependent terms proportional to the $\pi$-$N$ sigma
term at one-loop.  Further analysis is in progress [6].

{\bf References}\\
{}[1] I. Zahed and H. Yamagishi, Ann. Phys. 246, 3 (96).\\
{}[2] J.V. Steele, I. Zahed, and H. Yamagishi, hep-ph/9505330\\
{}[3] J.V. Steele, I. Zahed, and H. Yamagishi, hep-ph/9512233\\
{}[4] J. Gasser, M.E. Sainio, and A. \v{S}varc NPB307 (88) 779.\\
{}[5] V. Bernard, N. Kaiser, T.S.H. Lee, and Ulf-G. Mei{\ss}ner,
Phys. Rep. 246 (94) 315; V. Bernard, N. Kaiser, and
Ulf-G. Mei{\ss}ner, hep-ph/9507418.\\
{}[6] J.V. Steele, I. Zahed, and H. Yamagishi, in preparation.
\newpage 
\newcommand{\be}{\begin{eqnarray}}
\newcommand{\ee}{\end{eqnarray}}
\begin{center}
{\large\bf Scalar susceptibility and critical behavior in QCD and Schwinger
model}\\[0.5cm]
{\bf Andrei Smilga}$^{a)}$ and Jack Verbaarschot$^{(b)}$\\
a) ITEP, Moscow, Russia \\
b) SUNY at Stony Brook, Stony Brook, USA
\end{center}

We evaluate the leading infrared behavior of the scalar susceptibility
\be
\chi = \int d^4x \langle \sum_{i=1}^{N_f}\bar q_i q_i (x) \ \sum_{i=1}^{N_f}
\bar q_i q_i(0)
\rangle - V \langle \sum_{i=1}^{N_f}\bar q_i
q_i \rangle ^2 = \frac 1V \left . \partial^2_m \log Z  \right |_{m= 0},
\label{def}
\ee
in QCD and in the multiflavor Schwinger model for small non-zero fermion mass
$m$ and/or
small nonzero temperature as well as the scalar susceptibility for QCD at
finite volume.
In QCD, it is determined by one-loop chiral
perturbation theory, with the result that the leading infrared singularity
behaves as $\sim \log m$ at zero temperature:
\be
\chi^{IR} = \frac{N_f^2-1}{8\pi^2 }\left (\frac {\Sigma}{F_\pi^2} \right )^2
\log \frac{\Lambda^2}{M_\pi^2},
\label{chiQCD}
\ee
 and as $\sim T/\sqrt m$ at finite
temperature:
\be
\chi_T^{IR} = \frac{(N_f^2-1)}{ 4\pi} \frac T{\sqrt{2m}} \left (
\frac {\Sigma}{F_\pi^2}\right )^{3/2} \left( 1 +\frac 1{8N_f} \frac
{T^2}{F_\pi^2} \right).
\label{chiTQCD2lp}
\ee
(where also two loop chiral graphs are taken into account).
These are exact results to be checked in lattice and/or instanton model
numerical calculations.

In the Schwinger model with several flavors we use exact results for the
scalar correlation function. We find that the Schwinger model has a phase
transition at $T=0$ with critical exponents that satisfy the standard
scaling relations and do not coincide with the mean field theory predictions.
The singular behavior of this model depends on the
number of flavors with a scalar susceptibility that
behaves as $\sim m^{-2/(N_f+1)}$.
At finite volume $V$ we show that the scalar susceptibility is proportional
to $1/m^2V$. Recent lattice calculations of this quantity by Karsch
and Laermann [1] and the related lattice work by Kocic and Kogut [2]
are discussed.

{\bf References}\\
{}[1] F. Karsch and E. Laermann, Phys. Rev. {\bf D50} (1994) 6954. \\
{}[2] A. Kocic and J. Kogut, Phys. Rev. Lett. {\bf 74} (1995) 3109
\newpage 
\begin{center}
{\large\bf Finite Volume Analysis of Chiral Symmetry Breaking in QCD}\\[0.5cm]
{\bf Jan Stern}\\
\setcounter{footnote}{0}
Division de Physique Th\'eorique\footnote{Laboratoire de Recherche des
Universit\'es Paris XI et Paris VI, associ\'e au CNRS}, Institut de Physique
Nucl\'eaire\\
F-91406 Orsay Cedex, France
\end{center}
It is argued [1] that in QCD, there exists a natural
possibility of spontaneous breakdown of chiral symmetry (SBCHS) without
quark-antiquark condensation in the vacuum. In contrast to the Nambu
Jona-Lassinio model, the ground state of QCD might be characterized by a
non Gaussian distribution of small eigenvalues of the (Euclidean) Dirac
operator. This could lead to the large volume behaviour $V^{-1/2}$ of the
averaged lowest Dirac levels. The corresponding level density would not
be sufficient to trigger ${\bar q}q$ condensation [2,3]  ( the large volume
behaviour $V^{-1}$ is necessary ), although it could be sufficient to
make appear Goldstone bosons coupled to the conserved axial-vector currents.
New sum rules for inverse powers of the Dirac eigenvalues are derived [3,1]
which could be suitable for a numerical study of the mechanism of SBCHS
not suffering from the usual drawbacks of lattice simulation near the
chiral limit.

{\bf References}\\
{}[1] J. Stern, in preparation.\\
{}[2] T. Banks and A. Casher, Nucl. Phys. B168 (1980) 103.\\
{}[3] H. Leutwyler and A. V. Smilga, Phys. Rev. D46 (1992) 5607.
\newpage 
\begin{center}
{\large\bf Elastic $\pi \pi$ scattering to two loops}\\[0.5cm]
{\bf Gilberto Colangelo}\\
Institut f\"ur Theoretische Physik\\
Bern, Switzerland
\end{center}

I have presented the calculation of the $\pi \pi$ scattering amplitude
to two loops in Chiral Perturbation Theory. The calculation has been
done without any approximations and the result is given in analytical
form [1].

A thorough numerical analysis is still in progress. However, if we use
the current values for the $O(p^4)$ constants, and estimate with
resonance saturation the new $O(p^6)$ constants, we get small
corrections to the quantities of direct experimental interest, like
$a_0^0$, $a_0^0-a_0^2$ and $\delta_0^0-\delta_1^1$.
As an example, $a_0^0$, which at one loop is $0.201$ [2], becomes
$0.217$. The two loop calculation confirms that CHPT can yield very
sharp predictions for $\pi \pi$ scattering at low energy, as stressed
in [2]. A clear discrepancy with experimental data would then require
a significant revision of our picture of the vacuum structure of QCD.
As shown in [3], a value of $a_0^0$ much larger than the CHPT
predictions would be the signal for a quark--antiquark condensate much
smaller than what is usually assumed.

Finally I have compared the CHPT predictions to a recent lattice
calculation of the two $S$--wave scattering lengths [4]. Despite the
systematic effects, such as quenching, the agreement is quite
impressive. It would be interesting to improve the study of these
scattering lengths on the lattice to clarify whether the agreement is
accidental or not.

\vskip 3cm

{\bf References}\\
{}[1] J. Bijnens, G. Colangelo, G. Ecker, J. Gasser, M. Sainio,
Phys. Lett. B374 (1996) 210.\\
{}[2] J. Gasser and H. Leutwyler, Phys. Lett. 125B (1983) 325.\\
{}[3] M. Knecht, B. Moussallam, J. Stern and N.H. Fuchs,
Nucl. Phys. B457 (1995) 513.\\
{}[4] M. Fukugita, Y. Kuramashi, M. Okawa, H. Mino and A. Ukawa, Phys.
Rev. D52 (1995) 3003.
\newpage 
\begin{center}
{\large\bf Low Energy $\pi$$\pi$ Scattering to Two Loops}\\[0.5cm]
{\bf Marc Knecht}\\
\setcounter{footnote}{0}
Division de Physique Th\'eorique\footnote{Laboratoire de Recherche des
Universit\'es Paris XI et Paris VI, associ\'e au CNRS}, Institut de Physique
Nucl\'eaire\\
F-91406 Orsay Cedex, France
\end{center}
The amplitude for elastic $\pi$$\pi$ scattering at low energy has been
computed to two-loop accuracy in the chiral expansion [1].
As shown previously [2], it is determined by
the Goldstone nature of the pion, combined with the general S-matrix
properties of analyticity, crossing symmetry and unitarity, up to six
independent combinations of low energy constants, denoted by $\alpha$,
$\beta$, $\lambda_i$, $i=1,2,3,4$, and which
are not fixed by chiral symmetry. The four constants $\lambda_i$ were
determined via sum rules, evaluated using available data on $\pi$$\pi$
interaction at medium and high energies [3]. The remaining two parameters,
$\alpha$ and $\beta$, have to be determined from low energy $\pi$$\pi$ data.
An experimental determination of $\alpha$ is of particular importance. The
value of this parameter is intimately correlated to the ratio $x=-2{\widehat
m}\langle 0\vert {\bar q}q\vert 0\rangle /F_{\pi}^2M_{\pi}^2$, or
equivalently, to the quark mass ratio $m_s/{\widehat m}$. The commonly
accepted picture that spontaneous breakdown of chiral symmetry results from
a strong condensation of $q-{\bar q}$ pairs in the QCD vacuum requires that
$x\sim 1$ (or $r\sim 25$), and it is only compatible with values of $\alpha$
and $\beta$ close to unity, as predicted by standard $\chi$PT [4,5].
Unfortunately, a fit to the presently available $\pi$$\pi$ data obtained
from $K_{\ell 4}$ decays gives
$
\alpha = 2.16\pm 0.86,\ \beta = 1.074\pm 0.053
$,
and remains thus inconclusive in this respect.
Clearly, additional information, which would make such fits more accurate,
is needed. This may be provided by new high statistics $K_{\ell 4}$
experiments, from e. g. KLOE at DA$\Phi$NE, or by a precise measurement of
the lifetime of $\pi^+\pi^-$ atoms, as planed by the DIRAC experiment at CERN.

On the other hand, varying $\alpha$ and $\beta$ in the above ranges leads to
values of the threshold parameters which are in perfect agreement with the
results obtained from Roy equation analyses of available $K_{\ell 4}$ data.
In the range of energies accessible in $K_{\ell 4}$ decays,
the two-loop chiral expression of the $\pi\pi$ amplitude together with the
determination of the four constants $\lambda_i$ thus contains all the
relevant information on the $\pi\pi$ interaction which is already encoded in
the Roy equation.

{\bf References}\\
{}[1] M. Knecht, B. Moussallam, J. Stern and N. H. Fuchs, Nucl. Phys. B457
(1995) 513.\\
{}[2] J. Stern, H. Sazdjian and N. H. Fuchs, Phys. Rev. D47 (1993) 3814.\\
{}[3] M. Knecht, B. Moussallam, J. Stern and N. H. Fuchs, Nucl. Phys. B, in
print.\\
{}[4] J. Gasser and H. Leutwyler, Phys. Lett. B125 (1983) 325; Ann. Phys.
158 (1984) 142.\\
{}[5] J. Bijnens, G. Colangelo, G. Ecker, J. Gasser, M. Sainio,
Phys. Lett. B374 (1996) 210.
\newpage 
\begin{center}
{\large\bf  Constraints on $\pi$-$\pi$ scattering threshold
parameters from low energy sum rules}\\[0.5cm]
B. Ananthanarayan, {\bf D. Toublan} and G. Wanders\\
Institut de Physique Th\'eorique, Universit\'e de Lausanne,\\
CH 1015, Lausanne, Switzerland
\end{center}

In ChPT, the effective theory of QCD at low energy, the $\pi$-$\pi$
scattering threshold parameters play a central role [1,2]. Despite
the available phase shift analysis of the $S$- and $P$-waves, the
experimental information on the threshold parameters is not very
accurate [3,4].

$\pi$-$\pi$ scattering being a fundamental strong interaction
process well suited for theoretical investigations, the principles
of axiomatic field theory lead to a wealth of rigorous results [5].
Using analyticity, unitarity and crossing symmetry, and with the
help of  the homogeneous variables [6] three sum rules involving
dispersion integrals dominated by low-energy $S$- and $P$-waves can
be constructed from amplitudes which are completely symmetric in the
Mandelstam variables [7].

The dispersion integrals depend significantly on poorly known
threshold parameters.This lead us to a parametrization of the $S$-
and $P$-wave absorptive parts occuring in the integrands reproducing
the main features of the cross sections above threshold whereas the
scattering lengths and effective ranges remain free parameters [8].
The sum rules are turned into nonlinear equations for the $S$- and
$P$-wave threshold parameters and a combination of $D$-wave
scattering lengths. We show that the solutions of these equations
which are compatible with the data are confined to a rather small
portion of the experimentally allowed domain and enforce a strong
correlation between them [7]. This is our main result and it
establishes the relevance of our sum rules. Furthermore both
Standard and Generalized ChPT satisfy these sum rules constraints at
the one-loop level.

{\bf References}\\
{} [1] J. Gasser and H. Leutwyler, Ann. Phys. (N.Y.)158, 142 (1984).\\
{} [2] J. Stern, H. Sazdjian and N. H. Fuchs, Phys. Rev. D47, 3814
(1993).\\
{} [3] B. R. Martin, D. Morgan and G. Shaw,
``Pion-Pion Interaction in Particle Physics,'' Academic Press,
London/New York, 1976; also see W. Ochs, $\pi$ N News letter,
3, 25 (1991).\\
{} [4] M. M. Nagels, et al., Nucl. Phys. B147, 189 (1979).\\
{} [5] A. Martin, ``Scattering Theory:  Unitarity,
Analyticity and Crossing,'' Springer-Verlag, Berlin, Heidelberg,
New York, 1969.\\
{} [6] G. Wanders, Helv. Phys. Acta. 39, 228 (1966).\\
{} [7] B. Ananthanarayan, D. Toublan and G. Wanders, Phys. Rev.
D53, 2362 (1996).\\
{} [8] A. Schenk,  Nucl. Phys. B363, 97 (1991).
\newpage 
\centerline{\Large\bf Heavy Quark Solitons}
\baselineskip=18 true pt
\vskip 0.5cm
\centerline{{\bf Herbert Weigel}}
\vskip .2cm
\centerline{Institute for Theoretical Physics,
T\"ubingen University}
\centerline{Auf der Morgenstelle 14,
D-72076 T\"ubingen, Germany}
\vskip 0.7cm
\baselineskip=15pt

A generalization of the effective meson Lagrangian possessing the
heavy quark symmetry (HQS) [1] to finite meson masses is employed
to study the meson mass dependence of the spectrum of S-- and P
wave baryons containing one heavy quark or anti--quark. These
baryons are described as respectively heavy mesons or anti--mesons
bound in the background of a soliton of the light meson fields [2].
No further approximation is made to solve the bound state equations
for S-- and P wave heavy baryons. It is observed that the
HQS--prediction for the binding energies of these baryons
is approached only very slowly as the mass $M$ of the heavy meson
increases [3]. On the other hand the bound state wave--functions
satisfy the HQS--relations reasonably well at masses as low as
$M\approx5{\rm GeV}$ [3].

For physically relevant mass parameters associated with the charm
and bottom sectors, two types of models supporting soliton solutions
for the light mesons are considered: the Skyrme model of
pseudoscalars only as well as an extension containing also light
vector mesons. It has been found that only the Skyrme model with
vector mesons provides a reasonable description of the spectrum of
both light and heavy baryons [3]. It furthermore turns out that the
anti--quarks are unbound in the charm sector and only weakly bound,
if at all, in the bottom sector [3].  Subsequently the system
consisting of the vector meson soliton and the heavy meson bound
state is projected onto states with good spin and isospin. As
consistency check it has been shown that the mass gap between
heavy baryons with spin $\frac{1}{2}$ and $\frac{3}{2}$ decreases
as $1/M$. Turning again to the physical parameters the model
predicts the following mass differences:
$M(\Sigma_C)-M(\Lambda_C)=178{\rm MeV}$,
$M(\Lambda_C)-M(N)=1.321{\rm GeV}$ and
$M(\Lambda_B)-M(N)=4.495{\rm GeV}$. These compare reasonably
well with the empirical values $165{\rm MeV}$,
$1.345{\rm GeV}$ and $(4.701\pm0.050){\rm GeV}$, respectively.

In the heavy quark limit the coupling between mesons containing
a heavy quark and the soliton of the Nambu--Jona--Lasinio model is
studied in addition. As this soliton configuration contains quark
fields with non--vanishing grand spin conceptually different coupling
schemes between the heavy meson field and the soliton are discovered
[4]. These new schemes appear in addition to those which are already
present [2,3] in Skyrme type models and may yield a larger binding
of the baryon with a heavy quark [4].

{\bf References}\\
\noindent
{}[1] M.~Neubert,
Phys. Rep. {\bf 245} (1994) 259.\hfil\break
{}[2] C.~Callan and I.~Klebanov,
Nucl. Phys. {\bf B262} (1985) 365;\hfil\break
{}~~D.~P. Min, Y.Oh, B.~Park, and M.~Rho,
Int. J. Mod. Phys. {\bf E4} (1995) 47.\hfil\break
{}[3] J.~Schechter, A.~Subbaraman, S.~Vaidya, and H.~Weigel,
 Nucl. Phys. {\bf A590} (1995) 655;
E: Nucl. Phys. {\bf A598}, 583 (1996).\hfil\break
{}[4]L.~Gamberg, H.~Weigel, U.~Z\"uckert,
and H.~Reinhardt,
{\it Heavy Quark Solitons in the Nambu--Jona-Lasinio Model},
hep--ph/9512294.\hfil\break
\newpage 
\begin{center}
{\large\bf Low Energy QCD in the large $N_c$ Limit}\\[0.5cm]
{\bf Eduardo de Rafael}\\
 Centre de Physique Th\'eorique\\
         CNRS-Luminy, Case 907\\
         F-13288 Marseille Cedex 9, FRANCE.
\end{center}

My talk starts with a quick historical review of developments of QCD in
large $N_c$. I also compare the observed hadronic spectrum with the one
expected from large $N_c$. I use this as a motivation to introduce and
discuss the ENJL--model (see Ref.[1] and references therein,) as a low energy
model of QCD at large $N_c$ . The successes and drawbacks of this model are
reviewed.

I emphasize the need to develop good models for the low
energy QCD effective action. This is most dramatic in the non--leptonic
sector of the Standard Model. I show examples of low--energy
observables which require the knowledge of euclidean Green's functions at
all values of the euclidean momenta: the hadronic contributions to the muon
$g-2$; and the electromagnetic $\pi^{+}-\pi^{0}$ mass difference are two
examples I discuss in the light of the ENJL--model. I show how the problem
of matching long-- and short--distances can be successfully solved in these
two cases. I also discuss how the combined hypothesis: large $N_c$ and
$\langle\bar{\psi}\psi\rangle\rightarrow 0$, require at least two more
Weinberg sum rules which implies severe phenomenological constraints
(see Ref.[2].)

I next discuss, within the example of the Adler's function, the general
question of matching short distance QCD behavior obtained within the QCD
sum rules  \'{a} la SVZ, with the long distance hadronic behavior as
predicted by the ENJL--model. (See Ref.[3].) This clearly shows no overlap
of the two regimes, and the need for  a $G_{V}\ne 0$. I then discuss the
possibility of matching the two regimes using the QCD--Hadronic Duality
arguments developed in Ref.[4], and suggest a similar approach to solve the
problem of matching encountered in the recent calculations of the
light--by--light hadronic contributions to the muon $g-2$ and the
$B_{K}$--factor discussed in this Workshop, (see these proceedings.)

In the last part of my talk I review some recent exact results for low
energy observables  which, following Ref.[5], have been obtained within the
framework of a simultaneous expansion in the large
$N_c$ limit and $U(3)_{L}\times U(3)_{R}$ chiral perturbation theory. See in
particular Refs.[6] and [7].

{\bf References}\\
{}[1]~J. Bijnens, {\it Chiral Lagrangians and Nambu--Jona-Lasinio like
Models}, Phys. Reports 265, No6 (1996), and references therein.\\
{}[2]~J. Bijnens and E. de Rafael, {\it in progress.}\\
{}[3]~S. Peris, M. Perrottet and E. de Rafael, {\it in progress.}\\
{}[4]~R.A. Bertlmann, G. Launer and E. de Rafael, Nucl. Phys. B250, 61
(1985).\\
{}[5]~S. Peris and E. de Rafael, Phys. Letters, 348B, 539 (1995).\\
{}[6]~S. Peris, M. Perrottet and E. de Rafael, Phys. Letters 355B, 523
(1995).\\
{}[7]~A. Pich and E. de Rafael, hep-ph/9511465, to appear in Phys. Letters.
\newpage 
\begin{center}
{\large\bf Dispersive Analysis of the Predictions of}\\[0.5cm]
{\large{\bf Chiral Perturbation Theory for $\pi\pi$ Scattering}}\\[0.8cm]
{\bf {M.R. Pennington}}$\ ^1$ and J. Portol\'es$\ ^2$\\[0.1cm]
$^1$ Centre for Particle Theory, University of Durham, Durham DH1 3LE, U.K.\\
$^2$ INFN, Sezione di Napoli, I-80125 Naples, Italy
\end{center}
A way of testing the $\pi\pi$ predictions
of Chiral Perturbation Theory
against experimental data is to use dispersion relations
to continue experimental information into the subthreshold region
where the theory should unambiguously apply.  Chell and Olsson~[1]
have proposed a test of the
subthreshold behaviour of chiral expansions which highlights
 potential differences between the Standard~[2]
 and the Generalized~[3] forms of the theory.
We illustrate how, with current experimental uncertainties,
data cannot distinguish between
these particular {\it discriminatory} coefficients despite their
sensitivity~[4]. Nevertheless, the Chell-Olsson test does provide
a consistency check of the chiral expansion, requiring that the
${\cal O}(p^6)$ corrections to the {\it discriminatory} coefficients in
the Standard theory must be $\sim 100\%$. Indeed, some of these
 have been deduced~[5] from the
new ${\cal O}(p^6)$ computations~[6] and  found to give such large corrections.
One can then
check that the ${\cal O}(p^8)$ corrections must be much smaller.\\[2mm]
We conclude that this test, like others, cannot
distinguish between the different forms of Chiral Symmetry Breaking
embodied in the alternative versions of Chiral Perturbation Theory
without much more precise experimental information near threshold.\\[3mm]

{\bf References}\\
{}[1] E. Chell, Ph.D. thesis submitted to the University of Wisconsin;\\
M.G. Olsson, {\it Chiral Dynamics: Theory and Experiment},
 eds. A.M. Bernstein,\\
B.R. Holstein, (Springer-Verlag, 1995) pp. 111-112.\\
{}[2] J. Gasser and H. Leutwyler, Ann. Phys. (NY) 158 (1984) 142;
Nucl. Phys.\\
B250 (1985) 465.\\
{}[3] J. Stern, H. Sazdjian and N.H. Fuchs, Phys. Rev. D47 (1993) 3814.\\
{}[4] M.R. Pennington and J. Portol\'es, submitted to Physical Review D.\\
{}[5] B. Moussallam, private communication.\\
{}[6] M. Knecht, B. Moussallam, J. Stern and N.H. Fuchs, Nucl. Phys.
B457 (1995) 513;
J. Bijnens, G. Colangelo, G. Ecker, J. Gasser and M. Sainio,
Phys. Lett. B374 (1996) 210.
\newpage 
\def\beq{\begin{equation}}
\def\eeq{\end{equation}}
\def\bed{\begin{displaymath}}
\def\eed{\end{displaymath}}
\def\beqq{\begin{eqnarray}}
\def\eeqq{\end{eqnarray}}
\def\bedd{\begin{eqnarray*}}
\def\eedd{\end{eqnarray*}}
\begin{center}
{\large\bf FSI in $\eta\rightarrow 3\pi$ and the quark mass
ratio $Q^2$}\\[0.5cm]
{\bf Christian Wiesendanger}\\
Dublin Institute for Advanced Studies, School of Theoretical
Physics\\
10 Burlington Road, Dublin 4, Ireland\\
\end{center}

To leading order the mass ratios of the three light quark flavours
$u,d,s$ are easily accessible and known for a long time. The
next-to-leading order analysis has been performed by Gasser and
Leutwyler [1]. They have shown that the quantity
$Q^2=\frac{m_s^2-{\hat m}^2}{m_d^2-m_u^2}
=\frac{M_K^2}{M_{\pi}^2}
\frac{M_K^2-M_{\pi}^2}{M_{K^0}^2-M_{K^+}^2}
\left(1+O(m^2)\right)$
is given by the above ratio of pure QCD meson masses,
up to corrections of {\it second} order. To use the
experimental mass values for the mesons one has to
correct for the e.m. mass contributions. This is highly
controversial as Dashen's theorem may receive large
corrections [2].

An independent way to measure $Q^2$ is provided by the
isospin-violating
decay $\eta\rightarrow 3\pi$ as the corresponding rate is
proportional to $Q^{-4}$ [3]. Sutherland's theorem proves
to be stable [4]  and the main uncertainties in obtaining
a reliable rate come from the strong FSI of the $\pi$'s.
To evaluate those Kambor, Wiesendanger and Wyler [5] use extended
Khuri-Treiman equations. The subtraction to the dispersion relation
may then be fixed by the one-loop amplitude of Gasser and Leutwyler
[3]. The FSI corrections are moderate and enhance the amplitude
by 14\% at the center of the Dalitz plot. This reduces the usual
value for $Q^2=24.1$ obtained with Dashen to $Q^2=22.4\pm 0.9$. In agreement
with this result Anisovich and Leutwyler [6]
have obtained $Q^2=22.7\pm 0.8$ in their dispersive analysis.

{\bf References}\\
{}[1] J. Gasser and H. Leutwyler, Nucl. Phys. B250 (1985) 465.\\
{}[2] J. Donoghue, B. Holstein and D. Wyler, Phys. Rev.
D47 (1993) 2089;\\
R. Baur and R. Urech, Phys. Rev. D53 (1996) 6552;\\
J. Bijnens, Phys. Lett. B306 (1993) 343.\\
{}[3] J. Gasser and H. Leutwyler, Nucl. Phys. B250 (1985) 539.\\
{}[4] R. Baur, J. Kambor and D. Wyler, Nucl. Phys. B460 (1996) 127.\\
{}[5] J. Kambor, C. Wiesendanger and D. Wyler,
Nucl. Phys. B465 (1996) 215.\\
{}[6] A.V. Anisovich and H. Leutwyler, {\it Dispersive analysis
of the decay} $\eta\rightarrow 3\pi$, hep-ph/9601237.
\newpage 
\small
\vskip-1.5cm
\begin{center}
{\large\bf  Heavy Baryon ChPT with Light Deltas}\\
{\bf Thomas R. Hemmert}\\
Department of Physics and Astronomy, University of Massachusetts,
Amherst, MA 01003  USA
\end{center}
In recent years baryon chiral perturbation theory has matured into
a systematic field theory. Starting from work in the fully relativistic
framework [1], the introduction of the heavy mass formalism [2] led to the
development of the so called Heavy Baryon Chiral Perturbation Theory (HBChPT)
which allowed a consistent chiral power counting [3] to all orders.
Once one goes beyond the leading order lagrangian in the baryon sector, one
encounters two different classes of vertices: One class is accompanied
by unknown counterterms analogous to the the meson sector whereas
the other class corresponds to the so called 1/m corrections
(relativistic corrections) to vertices of lower order in the chiral expansion.
Both classes of vertices and the interplay between them are now well understood
for the case of spin 1/2 nucleons up to chiral order O($p^3$) [4].

In the scheme of HBChPT described so far, all baryon resonances are treated
as being infinitely heavy and decoupled from the theory [5]. Therefore they
only
contribute to higher order counterterms in the chiral lagrangian through
effective contact interactions. However, it is well known from phenomenology
that the first nucleon resonance ,$\Delta$(1232), plays a strong role in low
energy baryon processes. It has therefore been advocated for quite a while [6]
that one should keep the lowest lying spin 3/2 baryon resonances as explicit
degrees of freedom in the chiral lagrangian. Several calculations along this
line of thinking exist in the literature. However, many of these calculations
are incomplete. In particular, the construction of the above mentioned 1/m
corrected vertices involving spin 3/2 fields has been missing in the
literature.

We report on recent work [5,7] in SU(2) HBChPT that allows a systematic
treatment of spin 1/2 nucleon and explicit $\Delta$(1232) degrees of freedom.
Following the approach of [3], we start from the most general relativistic
spin 3/2 lagrangian, explicitly keeping ''point-transformation" invariance
and all possible ''off-shell" coupling structures. After having separated the
spin 3/2 and the spurious spin 1/2 components of the Rarita-Schwinger spinor
via a projector formalism, we make the transition to the heavy mass formalism.
To leading order, we reproduce the results of [3] (NN-sector) and [6]
($\Delta\Delta$,$\Delta N$-sector). In next-to-leading order [O($p^2$)], we
explicitly
construct all 1/m corrected vertices for the $NN$, $N\Delta$ and $\Delta\Delta$
lagrangians. We also discuss how the O($p^2$) $NN$ lagrangian of [3] has to be
changed, once one allows for explicit $\Delta$(1232) degress of freedom in the
theory. This leads us to a new understanding of ''resonance saturation" in the
baryon sector (see [5] for details). Furthermore, we discuss the O($p^2$)
vertices of the $\Delta\Delta$ and $N\Delta$ lagrangians accompanied by
counterterms and show how our methods can be generalised to obtain the
corresponding lagrangians beyond O($p^2$).

As a specific example we discuss the effect of $\Delta$(1232) in the process
of $\pi^0$ photoproduction at threshold. Keeping the delta-resonance in the
theory introduces a new mass scale $\Delta=M_{\Delta}-M_{N}\approx 300$MeV,
which is non-vanishing in the chiral limit, nevertheless small compared
with the chiral symmetry breaking scale $\Lambda_{\chi}\approx 1$GeV. We
therefore organise the calculation into a $\delta$-expansion, where $\delta$
corresponds to any of the small quantities $p,m_{\pi},\Delta$. Calculating
up to order $\delta^3$, we find that the leading order contribution is given
by a diagram involving one of the new O($p^2$) 1/m corrected
$N\Delta\pi$-vertices. We compare this result with a standard HBChPT
calculation [8] that has the deltas ''frozen out" and close with a numerical
estimate.\\
{}[1] J. Gasser, M.E. Sainio, A. Svarc; Nucl. Phys. {\bf B307} (1988) 779\\
{}[2] E. Jenkins, A. Manohar; Phys. Lett. {\bf B255} (1991) 558\\
{}[3] V. Bernard et al.; Nucl. Phys. {\bf B388} (1992) 315\\
{}[4] G. Ecker; Phys. Lett. {\bf B336} (1994) 508\\
{}[5] J. Kambor, these proceedings\\
{}[6] E. Jenkins, A. Manohar; Phys. Lett. {\bf B259} (1991) 353\\
{}[7] T.R. Hemmert, B.R. Holstein, J. Kambor; forthcoming\\
{}[8] V. Bernard, N. Kaiser, U.G. Meissner; Z.Phys. {\bf C70} (1996) 483
\newpage 
\normalsize
\begin{center}
{\large\bf Resonance Saturation in the Baryonic Sector of Chiral
Perturbation Theory}\\
\vspace {0.5cm}
{\bf Joachim Kambor}\\
Division de Physique Th\'eorique, Institut de Physique Nucl\'eaire\\
F-91406 Orsay Cedex
\end{center}
Heavy baryon chiral perturbation theory (HBChPT) [1,2] including spin 3/2
delta-resonance
degrees of freedom [3] has recently been reformulated by making use of a
1/m-expansion, m beeing the nucleon mass [4,5,6]. The theory admits a
systematic
expansion in the small scale $\delta$, where $\delta$ collectively denotes
soft momenta, the pion mass or the delta-nucleon mass difference. It is pointed
out that valuable information about HBChPT can be obtained by comparing the
chiral expansion with the $\delta$-expansion. Large corrections originating
from intermediate deltas can be identified and the convergence of the chiral
expansion with respect to these effects can be studied. Renormalization as well
as the different meaning of counterterms in HBChPT and the $\delta$-expansion,
respectively, is discussed in detail. This leads directly to a
reformulation of resonance saturation in the baryonic sector of ChPT [6].
As an explicit example, the scalar sector of one-nucleon processes in
chiral SU(2) is worked through. In particular, it is shown that the shift of
the scalar form factor of the nucleon between the Cheng-Dashen point and
zero, $\sigma(2 m_\pi^2)-\sigma(0) \approx 15$ MeV [7], has a natural
explanation in the $\delta$-expansion.

\vspace {0.5cm}

{\bf References}

\vspace{0.3cm}\noindent
{}[1] J. Gasser, M.E. Sainio, and A. Svarc,
Nucl. Phys. {\bf B307} (1988) 779. \\
{}[2] V. Bernard et al., Nucl. Phys. {\bf B388} (1992) 315.\\
{}[3] E. Jenkins and A.V. Manohar, Phys. Lett. {\bf B259} (1991) 353.\\
{}[4] T.R. Hemmert, these proceedings. \\
{}[5] J. Kambor, Heavy Baryon Chiral Perturbation Theory and the Spin
3/2 Delta Resonances, talk given at the 7th International Conference
on the Structure of Baryons, Santa Fe, NM, 3-7 Oct 1995,
Orsay preprint IPNO/TH 96-13. \\
{}[6] T.R. Hemmert, B.R. Holstein, and J. Kambor, in preparation. \\
{}[7] J. Gasser, H. Leutwyler, and M.E. Sainio, Phys. Lett. {\bf B253} (1991)
252, 260.
\newpage 
\begin{center}
{\large\bf Novel Algebraic Consequences of Chiral Symmetry}\\[0.5cm]
{\bf Silas R. Beane}\\
Duke University\\
Durham, NC 27708-0305\\
USA
\end{center}

The empirical success of the Ademollo-Veneziano-Weinberg mass
relation provides an example of regularity in the hadronic spectrum
that remains unexplained by the symmetries of QCD [1]. We provide an
explanation for the success of this relation based on the premise that
all hadrons fill out ---in general--- reducible representations of
$SU(2)\times SU(2)$ [2].  Mass-squared matrix elements of heavy hadrons
and light hadrons are related using heavy quark and chiral
symmetries [3]. Our result suggests that hadrons might be profitably
viewed as bound states of weakly interacting, parity-doubled
constituent quarks. We illustrate the essence of our result using a
simple effective lagrangian model.

{\bf References}\\
{}[1] M. Ademollo, G. Veneziano, and S. Weinberg,
  Phys. Rev. Lett. 28 (1968) 83.\\
{}[2] S. Weinberg,
  Phys. Rev. Lett. 65 (1990) 1177; {\it ibid}, 1181.\\
{}[3] S.R. Beane, DUKE-TH-95-98, hep-ph/9521228.
\newpage 
\small
\begin{center}
{\large\bf Hadronic contributions to the muon g-2: an updated analysis}
\\{\bf Elisabetta Pallante}, Johan Bijnens and Joaqu\'\i m Prades.\\
NORDITA, Blegdamsvej 17, DK-2100,
 Copenhagen, Denmark
\end{center}
The anomalous magnetic moment of the muon is one of the best candidates
to probe the electroweak sector of the Standard Model. For a review of
the theoretical aspects see e.g. [1].
Three types of contributions are present: the pure QED contributions,
the hadronic contributions and the weak contributions.
Pure QED contributions are
largely dominant $a_\mu^{\rm QED} = 11 \, 658 \, 470.6 (0.2) \cdot 10^{-10}$
[1],
but the leading hadronic vacuum
polarization contribution is sizable $a_\mu^{\rm h.v.p.} =
725.04 (15.76) \cdot 10^{-10}$ [2] and theoretically
predicted with an
uncertainty of the same size of the weak contributions
$a_\mu^{\rm EW}= 15.1 (0.4) \cdot 10^{-10}$ [3].
To disentangle weak contributions we need to further reduce the
theoretical uncertainty which affects the hadronic sector.
A new BNL experiment is planning to reach an
accuracy of $\pm 40\cdot 10^{-11}$ in the determination of the muon
$a_\mu =(g-2)/2$,
more than a factor of twenty of reduction respect to the latest
determination at the Cern Storage Ring
$a_\mu^{\rm exp} = 11 \, 659 \, 230 (84) \cdot 10^{-10}$.
 This motivated the recently
raised interest in the theoretical
determination of this observable with an improved accuracy.
At present hadronic contributions are the main source of uncertainty in
the theoretical prediction.
We distinguish three classes of hadronic contributions to $a_\mu$:
a) hadronic vacuum polarization (h.v.p.) contribution which appears at order
$(\alpha/\pi )^2$ b) higher order corrections to the hadronic vacuum
polarization diagram [4] and c) light-by-light scattering contributions
which start at order $(\alpha/\pi )^3$.

The h.v.p. contribution can be extracted via phenomenological dispersive
analysis from the total $e^+e^-\to hadrons$ cross section. This is at
present the most accurate way of determination [2]. To further reduce its
uncertainty new more precise
data of $e^+e^-\to hadrons$ cross section are needed.
Alternatively a low energy effective model of QCD can be used.
In spite of the lack of confinement and the theoretical debated
connection with QCD the Extended Nambu-Jona Lasinio (ENJL)
model [5] does satisfy few phenomenological
constraints (e.g. Weinberg Sum Rules) which are necessary conditions
to guarantee a good matching with PQCD and provides a systematic
treatment of observables dominated by long distance dynamics.
Its prediction of the h.v.p. contribution [6] is in good agreement with
phenomenological determinations.

A novel determination of the light-by-light scattering contribution
has been proposed in [7] within the ENJL framework
(see also [8] for an alternative derivation).
The dominant contribution
is the twice anomalous pseudoscalar exchange diagram.
The final result we get is $a_\mu^{\rm light-by-light}=
 (-9.2\pm3.2 )  \cdot 10^{-10}$.
This is between two and three times the expected experimental uncertainty
at the forthcoming BNL muon $g-2$ experiment.
Adding the other
Standard Model contributions to $a_\mu$ the
present theoretical estimate for the muon $g-2$ is
$a^{\rm th}_\mu= 11\, 659 \, 182 (16)  \cdot 10^{-10}$.\\
{}[1]
``Frontiers of High Energy Spin
   Physics'',Nagoya
1992,
    T. Hasegawa et al. (eds.), (Universal Acad. Press, Tokyo, 1992);
    ``The Future of Muon Physics'',
    Heidelberg,
Germany (1991), Z. Phys. C56, K. Jungmann,
   V.W. Hughes, and G. zu
Putliz (eds.);
``Quantum Electrodynamics'', T. Kinoshita (ed.),
   World Scientific, Singapore,
 (1990).       \\
{}[2] S. Eidelman and F. Jegerlehner, Z. Phys. C67 (1995) 585 .\\
{}[3] A. Czarnecki, B. Krause, and W.J. Marciano, Phys. Rev.
      D52(1995)2619,
     Karlsruhe preprint
TTP95-34(1995), hep-ph/9512369;
S. Peris, M. Perrottet, and E. de Rafael, Phys. Let.
       B355(1995)523.\\
{}[4] T. Kinoshita, B. Ni{\u{z}}i\'c, and Y. Okamoto,
      Phys. Rev. D31 (1985) 2108.\\
{}[5] J. Bijnens, C. Bruno and E. de Rafael, Nucl. Phys.
      B390(1993)501;
J. Bijnens, Phys. Rep. 265(1996)369;\\
{}[6] E. de Rafael, Phys. Lett. B322 (1994) 239;
E. Pallante, Phys. Lett. B341 (1994) 221.\\
{}[7] J. Bijnens, E. Pallante, and J. Prades,
      Phys. Rev. Lett. 75 (1995) 1447;
Erratum: ibid. 75 (1995) 3781;
 hep-ph/9511388, to appear in Nucl. Phys. B.\\
{}[8] M. Hayakawa, T. Kinoshita, and A.I. Sanda,
      Phys. Rev. Lett. 75 (1995) 790;
Nagoya Univ. preprint DPNU-95-30 (1995).
\newpage 
\normalsize
\begin{center}
{\large\bf Some Hadronic Matrix Elements within the
Extended NJL Model}\\[0.5cm]
Johan Bijnens$^{a)}$ and {\bf Joaquim Prades}$^{b)}$\\[0.5cm]
$^a$NORDITA, Blegdamsvej 17, DK-2100 Copenhagen (Denmark).\\[0.3cm]
$^b$Departament de F\'{\i}sica Te\`orica, Universitat de Val\`encia
and IFIC, Universitat de Val\`encia-CSIC\\
C/ del Dr. Moliner 50, E-46100 Burjassot (Val\`encia) Spain.
\end{center}

\vspace{0.5cm}
The test of the Standard Model at low energy is generally
affected by the uncertainty associated with the calculation
of hadronic matrix elements at low energy.
Even at low energies, the calculation of hadronic matrix elements
requires, in general, the knowledge of the strong interacions at
all scales. This is for instance what happens in the $\hat B_K$
parameter [1] or in the corrections to the Dashen's theorem [2].
There, a virtual boson ($W$s or photon) is integrated out making
the internal scale to run from zero up to $M_W$ ($\infty$).
There are other hadronic matrix elements (like $\gamma \gamma
\to \pi^0 \pi^0$) [3] that start at high order within CHPT and
 therefore are more sensible to the high energy behaviour of QCD.
Also in this type of hadronic matrix elements one would like
to obtain some matching with QCD.
We have atacked the problem of calculating hadronic matrix
elements using the Extended NJL model version
in Refs. [4,5,6,7] as a good hadronic model at low energies
and imposing short distance QCD
behaviour at high energies. Though the matching obtained is not very
good and more work to improve the intermediate energy region
is needed, we have already obtained interesting results [1,2,3].
Work in the same direction is in progress for $\Delta S = 1$ decays
like $K \to \pi, 2\pi$. For some work to improve the matching between
the low-energy contributions and the shorty distance for two
point functions, see the contribution
by Eduardo de Rafael to this Workshop and references therein.

\vspace{0.5cm}

{\bf References}\\
{}[1] J. Bijnens and J. Prades, Nucl. Phys. B444 (1995) 523.\\
{}[2] J. Bijnens and J. Prades, in preparation.\\
{}[3] J. Bijnens, A. Fayyazuddin, and. J. Prades, preprint
     FTUV/95-70 (1995) (to be published in Phys. Lett. B).\\
{}[4] J. Bijnens, C. Bruno, and E. de Rafael, Nucl. Phys. B390 (1993) 501.\\
{}[5] J. Bijnens, E. de Rafael, and H. Zheng Z. Phys. C62 (1994) 437.\\
{}[6] J. Bijnens and J. Prades, Z. Phys. C64 (1994) 475.\\
{}[7] J. Bijnens, Phys. Rep. 265 (1996) 369.
\newpage 
\begin{center}
{\large\bf On the Corrections to Dashen's Theorem}\\[0.5cm]
{\bf Res Urech}\\
Institut f\"ur Theoretische Teilchenphysik, Universit\"at Karlsruhe\\
D-76128 Karlsruhe, Germany
\end{center}
The electromagnetic corrections to the masses of the pseudoscalar
mesons $\pi$ and $K$ are considered. At order $O(e^2)$ in the chiral limit
Dashen's theorem [1] is given by the relation $
\Delta M^2_K - \Delta M^2_\pi = 0$, where $\Delta M^2_P = M^2_{P^\pm} -
M^2_{P^0}$. At order $O(e^2 m_q)$ this relation is subject to corrections,
which are probably large [2]. We calculate the contributions at order
$O(e^2 m_q)$ that arise from resonances within a photon loop in the
framework of chiral perturbation theory [3]. Within this approach we find
rather moderate deviations to Dashen's theorem [4].\\[0.5cm]

{\bf References}\\[0.2cm]
\begin{tabular}{ll}
{}[1]& R.Dashen, Phys. Rev. 183 (1969) 1245.\\
{}[2]& K.Maltman and D.Kotchan, Mod. Phys. Lett. A5 (1990) 2457;\\
 &     J.F.Donoghue, B.R.Holstein and D.Wyler, Phys. Rev. D47 (1993) 2089;\\
 &     J.Bijnens, Phys. Lett. B306 (1993) 343;\\
 &     R.Urech, Nucl. Phys. B433 (1995) 234;\\
 &     H.Neufeld and H.Rupertsberger, Z. Phys. C68 (1995) 91.\\
{}[3]& G.Ecker, J.Gasser, A.Pich and E.de Rafael, Nucl. Phys. B321
      (1989) 311.\\
{}[4]& R.Baur and R.Urech, Phys. Rev. D53 (1996) 6552.
\end{tabular}
\newpage 
\begin{center}
{\large\bf $K \rightarrow \pi \gamma \gamma$ decays~:
unitarity corrections}\\[0.2cm]
{\large \bf and vector meson contributions}\\[0.5cm]
{\bf G. D'Ambrosio} and J. Portol\'es\\[0.1cm]
INFN, Sezione di Napoli, I-80125 Naples, Italy
\end{center}
\vspace*{0.1cm}
$K \rightarrow \pi \gamma \gamma$ are interesting processes by
themselves as ChPT tests and $K_L \rightarrow \pi^{\circ} \gamma
\gamma$ in particular might have an important r$\hat{o}$le as a CP conserving
amplitude contributing to $K_L \rightarrow \pi^{\circ} e^+ e^-$.
\par
Two different helicity amplitudes contribute to $K_L \rightarrow
\pi^{\circ} \gamma \gamma$~: $A$ and $B$. The first appears at
${\cal O}(p^4)$~[1], it is vanishing for small diphoton invariant mass
and generates a suppressed amplitude for $K_L \rightarrow \pi^{\circ}
e^+ e^-$. The second amplitude $B$ appears at ${\cal O}(p^6)$, it is
non--vanishing for small diphoton invariant mass and generates an
unsuppressed amplitude for $K_L \rightarrow \pi^{\circ} e^+ e^-$~[2].
Though the experimental spectrum for $K_L \rightarrow \pi^{\circ}
\gamma \gamma$ seems very well reproduced by the ${\cal O}(p^4)$
leading contribution the rate is not. This has lead several authors
to consider some ${\cal O}(p^6)$ contributions (see references quoted
in [2])~: i) unitarity corrections
from physical $K_L \rightarrow \pi^{\circ} \pi^+ \pi^-$ amplitude give
a $20-30 \%$ increase in the amplitude with a slight deformation of the
${\cal O}(p^4)$ spectrum, ii) an appropriate choice of the $B$ amplitude
generated by vector meson contributions can accomodate width and
spectrum, and also iii) unitarization of the $\pi \pi$ intermediate
states amplitude  with inclusion of the
experimental $\gamma \gamma \rightarrow \pi^{\circ} \pi^{\circ}$ amplitude
should  help.
\par
$K^+ \rightarrow \pi^+ \gamma \gamma$ is also an appealing channel which
will be measured soon. The leading contribution is ${\cal O}(p^4)$ with
loops and local contributions which size is an interesting test of weak
hadron dynamics [3].
\par
We show [4] that unitarity corrections to $K^+ \rightarrow \pi^+ \gamma
\gamma$ are important and generate also a $20-30 \%$ increase in
the $B$ amplitude. We then study [5] ${\cal O}(p^6)$ vector meson models
contributing to this channel and to $K_L \rightarrow \pi^{\circ}
\gamma \gamma$ showing that local contributions generated by
vector meson exchange in the charged channel are likely to be
negligible contrarily to the neutral channel. This can be studied in the
spectrum for small diphoton invariant mass, while the ${\cal O}(p^4)$
unknown local contribution can be determined from the rest of the
kinematical region, or from the rate.
\vspace*{0.3cm} \\

{\bf References} \\
{}[1] G. Ecker, A. Pich, E. de Rafael, {\em Phys. Lett.}, {\bf B189}
(1987) 363. \\
$\; \; \; \; \; $ L. Cappiello, G. D'Ambrosio, {\em Nuovo Cimento},
{\bf 99A} (1988) 155. \\
{}[2] G. D'Ambrosio, G. Ecker, G. Isidori, H. Neufeld, ``Radiative
non--leptonic kaon decays" in the Second DA$\Phi$NE Physics Handbook,
ed. by L. Maiani, G. Pancheri, N. Paver, LNF (1995), p.265. \\
{}[3] G. Ecker, A. Pich, E. de Rafael, {\em Nucl. Phys.}, {\bf B303}
(1988) 665. \\
{}[4] G. D'Ambrosio, J. Portol\'es, Preprint INFNNA--IV--96/12,
hep-ph 9606213.\\
{}[5] G. D'Ambrosio, J. Portol\'es, Preprint INFNNA--IV--96/21.
\newpage 
\begin{center}
{\large\bf  Radiative Four--Meson Amplitudes in CHPT} \\[0.5 cm]
{\bf Gino Isidori} \\
INFN, Laboratori Nazionali di Frascati,
P.O. Box 13, I--00044 Frascati, Italy
\end{center}

Chiral perturbation theory (CHPT) [1,2,3] naturally
incorporates electromagnetic gauge invariance. To lowest order in the
derivative expansion, $O(p^2)$ in the meson sector, amplitudes
for radiative transitions are completely determined by the corresponding
non--radiative amplitudes. Direct emission (DE),
carrying genuinely new information, appears only at
$O(p^4)$. In the case of $K \to 2\pi\gamma$ and $K \to 3\pi\gamma$ decays,
the study of these information is of great interest
to understand the structure of the $O(p^4)$ nonleptonic weak
lagrangian [4,5].

The fact that $O(p^2)$ radiative amplitudes are
completely determined by the corresponding
non--radiative ones is a consequence of  Low's theorem [6].
In the case of radiative three--meson processes,
like $K\to 2 \pi\gamma$ decays, where the on--shell non--radiative
amplitude is constant, it is straightforward to extend the relation between
radiative and non--radiative amplitudes  to higher orders
in the chiral expansion [7,8,9,10]. On the other hand,
in the case of radiative four--meson processes, the
dependence  from kinematical variables  of the non--radiative amplitudes
makes this extension less trivial.
It has been shown in Ref. [11] how to extend  Low's theorem
by means of second derivatives of the non--radiative amplitudes
to define a ``generalized bremsstrahlung'' (GB).
This amplitude include all the contributions to the radiative process
generated by local $O(p^4)$ counterterms that contribute to the
non--radiative one. By this way, the remaining part of the
radiative amplitude receives $O(p^4)$ contributions only form genuine
radiative counterterms
(operators with an explicit electromagnetic strength tensor)
and from loop diagrams.

In principle, the GB can be calculated using the experimental
information on the non radiative process, minimizing the
uncertainties related to higher order effects in CHPT. On the other hand,
the remaining contributions must be computed using
$O(p^4)$ CHPT predictions.
The only loop diagrams that contribute to the DE,
i.e. which are not included in the GB, are the so--called ``fish--diagrams''.
In Ref. [11] a compact but completely general
expression for these loop amplitudes has been presented.
Using the most general parametrization of the $O(p^2)$
four--meson vertices, the loop amplitudes of Ref. [11]
can be applied to any radiative
four--meson process, both in the
strong and in the weak sector (known results for
$K\to 2\pi \gamma$ decays [7,8,9,10] are recovered as a particular case).

Detailed numerical analysis for  $K\to 3\pi \gamma$ and
$\eta\to 3\pi \gamma$ transitions are in progress [12].

{\bf References}\\
{}[1] S. Weinberg, Physica 96A (1979) 327.\\
{}[2] J. Gasser and H. Leutwyler, Ann. Phys. 158 (1984) 142. \\
{}[3] J. Gasser and H. Leutwyler, Nucl. Phys. B250 (1985) 465. \\
{}[4] J. Kambor, J. Missimer and D. Wyler, Nucl. Phys. B346 (1990) 17. \\
{}[5] G. Ecker, J. Kambor and D. Wyler, Nucl. Phys. B394 (1993) 101. \\
{}[6] F.E. Low, Phys. Rev. 110 (1958) 974. \\
{}[7] G. Ecker, H. Neufeld and A. Pich, Phys. Lett. B278 (1992) 337. \\
{}[8] G. D'Ambrosio, M. Miragliuolo and F. Sannino, Z. Phys. C59 (1993) 451. \\
{}[9] G. Ecker, H. Neufeld and A. Pich, Nucl. Phys. B413 (1994) 321. \\
{}[10] G. D'Ambrosio and G. Isidori, Z. Phys. C65 (1995) 649. \\
{}[11] G. D'Ambrosio, G. Ecker, G. Isidori and H. Neufeld, Preprint
  INFNNA-IV-96/11 [hep-ph/9603345], to appear in Phys. Lett. B\\
{}[12] G. D'Ambrosio, G. Ecker, G. Isidori and H. Neufeld, in preparation.
\newpage 
\begin{center}
{\large \bf Aspects of Renormalization in Chiral Perturbation Theory}
\\[0.5cm]
{\bf Gerhard Ecker}\\
Inst. Theor. Physik, Univ. Wien, Vienna, Austria\\[0.3cm]
\end{center}

In a short introduction to the loop expansion in CHPT, I discussed
the advantages for a consistent chiral power counting of using
the lowest--order mesonic chiral Lagrangian of $O(p^2)$
to define the classical solution as the starting point for the loop
expansion to any chiral order. One important implication is that
the lowest--order equation of motion can be used in the generating
functional for any chiral order.
The renormalization procedure for the $\pi\pi$ scattering amplitude
as well as for $M_\pi$ and $F_\pi$ to $O(p^6)$ [1] was the main topic
of my presentation at the Workshop. Two--loop, one--loop and tree--level
diagrams add up to the final renormalized quantities. Various consistency
checks for the calculation
were discussed that are essentially due to the proper handling of
subdivergences of $O(p^4)$. One of these conditions allows for the calculation
of the leading squares of chiral logs appearing at $O(p^6)$ in terms of
one--loop diagrams only (with a single $O(p^4)$ vertex) [2,3].
In the last part, I analysed the mesonic generating functional of $O(p^4)$
with one off--shell meson line to arrive at the following general conclusions:
\begin{enumerate}
\item In the calculation of the meson--baryon functional of $O(p^3)$
[4,5] in heavy--baryon CHPT, the relative contributions of the local
meson--baryon action and of the reducible tree--level diagrams with one vertex
from the mesonic Lagrangian of $O(p^4)$ depend on the choice of the
latter Lagrangian, i.e. on the choice of meson fields. The sum is
of course independent of the chosen convention.
\item Expanding the mesonic low--energy constants of $O(p^4)$ in a
Laurent expansion around $d=4$, the coefficients linear in $d-4$ appear in
general in mesonic amplitudes of $O(p^6)$ due to two--loop diagrams. These
terms can always be absorbed in the coupling constants of $O(p^6)$
in a process independent fashion. In other words, those coefficients are
not measurable quantities independent of the low--energy constants
of $O(p^6)$.
\end{enumerate}

{\bf References}\\
{}[1] J. Bijnens, G. Colangelo, G. Ecker, J. Gasser and M.E. Sainio,
Phys. Lett. B374 (1996) 210;
more detailed version in preparation.\\
{}[2] S. Weinberg, Physica 96A (1979) 327.\\
{}[3] G. Colangelo, Phys. Lett. B350 (1995) 85; ibid. B361 (1995) 234 (E).\\
{}[4] G. Ecker, Phys. Lett. B336 (1994) 508.\\
{}[5] G. Ecker and M. Moj\v zi\v s, Phys. Lett. B365 (1996) 312.
\newpage 
\begin{center}
{\large\bf Chiral Symmetry and Hypernuclei}\\[0.5cm]
{\bf Roxanne P. Springer}\\
Duke University\\
Durham, North Carolina, USA
\end{center}
Hypernuclei provide another laboratory for testing
predictions of heavy baryon chiral perturbation theory[1].
In large-A nuclei, the free $\Lambda \rightarrow {\rm N} \pi$
mesonic decay is Pauli blocked.  Instead, the hypernucleus decays through
the (nonmesonic) reaction $\Lambda {\rm N} \rightarrow {\rm NN}$. The meson
exchange contribution to this process is dominated by pion
exchange, where both the weak and strong vertices required can
be found experimentally.
The total nonmesonic decay widths are well reproduced using
a variety of models, while the ratio of proton induced
to neutron induced decays is much more difficult to understand.
Shell model calculations on $^{12}_{\Lambda}C$[2] indicate
that the kaon meson exchange plays an important role in
such ratios.  We calculate the leading SU(3) breaking
one-loop corrections to the weak KNN couplings relevant
for this decay [3]. One of the motivations for this calculation is to further
investigate the hyperon p-wave problem.  For many years
it has been known that the chiral coefficients dictated
by the s-wave hyperon decays reproduce the p-wave
data very poorly.  Further, the leading
logarithmic loop calculations for this process are
found to be large, yet still in severe disagreement
with the data [4,5].  This finding led to concerns about
the validity of chiral perturbation theory for this
process [6].  The p-wave KNN couplings that we calculate arise
from the same set of diagrams which correct the p-wave
hyperon decays.  Therefore, a comparison of this calculation
with data extracted from hypernuclear decays may lead to
a better understanding of what should be expected from chiral
perturbation theory in this sector.
We find that the leading logarithmic corrections to KNN
couplings are well behaved.  This supports the suggestion
that the problem in hyperon p-wave predictions comes from
accidental cancellations of tree level diagrams rather
than problems inherent in the theory[5].  The values for
the weak KNN couplings that we find are smaller than the
tree-level values.  These couplings will now be used
in a shell model calculation to test agreement with
experimental obervables[7].

{\bf References}\\
{}[1] E.Jenkins and A.Manohar, "Baryon Chiral Perturbation Theory,''
presented at Hungary, August, 1991.\\
{}[2] C.Bennhold, A.Parreno, A.Ramos, Few-Body Systems Suppl. (1996) 1.\\
{}[3] M.J.Savage and R.P.Springer, Phys. Rev. C53 (1996) 441.\\
{}[4] J.Bijnens, H.Sonoda, and M.B.Wise, Nucl. Phys. B261 (1985) 185.\\
{}[5] E.Jenkins, Nucl. Phys. B375 (1992) 561.\\
{}[6] C.Carone and H.Georgi, Nucl. Phys. B375 (1992) 243.\\
{}[7] C.Bennhold, private communication.
\newpage 
\setcounter{footnote}{0}
\renewcommand{\thefootnote}{\fnsymbol{footnote}}
\begin{center}
{\large\bf Hyperon Electromagnetic Properties in a Soliton Model}
\\[0.5cm]
{\bf Norberto N. Scoccola}\footnote[2]{On leave from Physics Dept.,
CNEA, Argentina and Fellow of the CONICET, Argentina.}\\
INFN, Sezione di Milano, via Celoria 16, I-20133 Milano, Italy.
\end{center}
The predictions obtained within the bound state soliton model[1]
for the electromagnetic decay widths of the decuplet
hyperons, the electromagnetic decay widths of the $\Lambda(1405)$
resonance and the electric and magnetic static polarizabilities of
the octet hyperons are discussed.
Details of this work are given in Refs.[2].

Our results for the radiative decay widths of the decuplet hyperons
are in good agreement with those obtained using the non-relativistic
quark model (NRQM), the bag model, heavy baryon chiral perturbation
theory (HBChPT) and quenched lattice QCD. This overall agreement between
different models contrasts with the situation for the $\Lambda(1405)$
decay widths. There, our predictions agree rather well with
the results of the cloudy bag model but are, however, much smaller
than those of the NRQM.  Concerning the static electric
polarizabilities we obtain rather small splittings between the values
corresponding to the different hyperons. Moreover, they are dominated
by the seagull terms which are basically given by the non-strange
contributions.
The structure is richer in the magnetic case because of the interplay
between a large (negative) seagull part with the relevant dispersive
contribution. Although some of our results are in agreement with those
of the NRQM,
in general this is not the case. In addition, the calculations
performed in the framework of HBChPT
lead to still different predictions. In this situation, it
is clear that the future experimental data from CEBAF and FNAL could
be of great help to discriminate among the different existing models
of hyperons.

{\bf References}\\
{}[1] C.G. Callan and I. Klebanov,
Nucl. Phys. {\bf B262} (1985) 365;
N.N. Scoccola, H. Nadeau, M.A. Nowak and M. Rho,
Phys. Lett. {\bf B201} (1988) 425;
C.G. Callan, K. Hornbostel and I. Klebanov,
Phys. Lett. {\bf B202} (1988) 269;
U. Blom, K. Dannbom and D.O. Riska,
Nucl. Phys. {\bf A493} (1989) 384.\\
{}[2]
C.L. Schat, C. Gobbi and N.N. Scoccola,
Phys. Lett. {\bf B356} (1995) 1;
C.L. Schat, N.N. Scoccola and C. Gobbi,
Nucl. Phys. {\bf A585} (1995) 627;
C. Gobbi, C.L. Schat and N.N. Scoccola,
Nucl. Phys. {\bf A598} (1996) 318.
\newpage 
\begin{center}
{\large\bf Strange contents in the nucleon \\
-- The effects of kaonic cloud -- \\
-- on the neutron electric form factor --}\\[0.5cm]
{\bf Teruaki Watabe} and Klaus Goeke\\
Institute for Theoretical Physics II,
Ruhr-University Bochum, \\
D-44780 Bochum, Federal Republic Germany
\end{center}

As well known the nucleon is constructed by three quarks.
For instance, the neutron has one up quark, which has a charge of
$+\frac{2}{3}$, and two down quarks, with a charge of $-\frac{1}{3}$.
The experiments indicate that the nucleon has an abundant structure measuring
its form factor and magnetic moment.
The electric charge distribution of the nucleon comes from the bare
valence quarks $(qqq)$ and valence quarks with the mesonic excitation of the
vacuum $(qqq^\prime + \bar{q}^\prime q)$.
The total charge of neutron is zero, therefore the bare valence quark
contribution to the electric charge distribution is nearly zero and the
neutron electric charge distribution is dominated by the mesonic excitation
of the vacuum.
Hence one can expect that mesonic clouds play quite important role on the
electric properties of the neutron.
We investigate them employing the chiral quark soliton model.
The chiral quark soliton ($\chi$QS) model provides a well-defined and reliable
framework in studying effects of mesonic clouds on the nucleon properties.
The $\chi$QS model is derived based on the instanton picture of the QCD
vacuum~[1] and described by a very simple QCD effective action, in
which quarks interact via Goldstone bosons.
The mesonic clouds in the $\chi$QS model are quite distinguished
from other hedgehog models, since they are generated by the Dirac-sea quark
polarization via one-quark loops.

The naive evaluation of the neutron electric form factor with the hedgehog
pion which has the Yukawa tail behavior characterized by the pion mass gives
a serious underestimation, because the kaon field arises as the rotational
excitation of the hedgehog pion field which has the same tail behavior as the
pion field.
We have solved this problem using the {\it hybrid method} of treating the
mesonic clouds~[2].
The neutron electric form factor is quite sensitive to the mesonic clouds and
the result of hybrid method fairly agrees with the experiments.
We have shown also the strange electric form factor and the square radius
using the hybrid method and obtained remarkably smaller results than those
appearing in the previous works done in the same model framework.
We have investigated also the proton electric properties, however they are
rather insensitive to the mesonic clouds.

{\bf References}\\
{}[1] D.I. Diakonov and V.Yu. Petrov,
Nucl.Phys. {\bf B272} (1986) 457.\\
{}[2] T. Watabe, H.-C. Kim and K. Goeke, preprint; {\sc rub-tpii-17/95},
e-print archive; hep-ph/9507318 (revised in May.1996), (1996).
\newpage 
\begin{center}
{\large\bf $\chi$PT description of the MSM:
 One and two loop
order}\\[0.2cm]
{\bf Joaquim Matias}\\
{\it Dip. di Fisica, Universita di Padova, Via F.Marzolo 8, I-35131
Padova} \end{center}

$\chi$PT provide us with a general parametrization,
 in terms of the coefficients of the
chiral lagrangian,
 of the symmetry
breaking sector of the SM at low energies.
 The coefficients of the operators compatibles with the symmetries[1]
(usually called chiral coefficients) can be fixed either from
the experiment, in that case we will end up with a model independent
description for the dynamics of the light fields ($W$ and $Z$), or
from the different possible candidates as underlying theory.
In the latter case the chiral coefficients will encode the
non-decoupling effects[2] of the heavy particle/s that have been
integrated out in each particular model. Due to the good agreement of the LEP
data with the MSM it seems reasonable to start by deriving the values
of the chiral coefficients $a_{i}^{\rm MSM}$[3]
corresponding to this model.
 The difference often minute between these $a_{i}^{\rm MSM}$ and the
contribution to the
chiral coefficients corresponding to the other alternative
models is where the clues of what lies beyond the SM are.
We consider a scenario with a Higgs large enough
to allow for a mass gap in between the light and heavy particle but
sufficiently light to be able to define a perturbative series.
The technique used to derive the chiral coefficients are the matching
conditions between
transverse connected Green functions. They have been extended to
include
the two next-to-leading corrections[4], the one-loop $1/M_{H}^{2}$ order
and the two-loop $M_{H}^{2}$ contribution, to the LEP1 relevant
coefficients. It is proposed a new formulation of the
matching conditions at higher loop orders that solves automatically the
subtleties
concerning gauge invariance and gives information on the scheme
dependence of the chiral coefficients.
As an outcome of the computation[4,5], it is shown how $\chi$PT
combined with the
properties of the on-shell scheme and the screening theorem of Veltman
provide us with a powerful tool. It allows to improve the usual
power
counting estimation of the Higgs contribution[6] to the renormalized
self-energies at nth-loop order by one power of $M_{H}^{2}$ less in the
$Z$ and $W$ renormalized self-energies and two powers less in the photon
self-energy. $\Delta \rho$, $\Delta r$ and $\Delta \kappa$ are
also computed. All conclusions can be made extensible to any other
perturbative underlying theory (MSSM, multiHiggs models, new gauge
extensions, \ldots).

{\bf References}\\
\noindent {}[1] A.Longhitano, Nucl. Phys. B188 (1981) 118.\\
{}[2] T.Appelquist and J.Carazzone, Phys. Rev. D11 (1975) 2856.\\
{}[3] M.J.Herrero and E.Ruiz Morales, Nucl. Phys. B418 (1994) 431;
Nucl. Phys. B437 (1995) 332; D.Espriu and J.Matias, Phys. Lett. B341
(1995) 332.\\
{}[4] J.Matias, Padova Preprint DFPD 96/TH/18, hep-ph 9604390.\\
{}[5] J.Matias and A.Vicini, in preparation.\\
{}[6] M.B.Einhorn and J.Wudka, Phys. Rev. D39 (1989) 2758.
\newpage 
\begin{center}
{\large\bf The Covariant Derivative Expansion Method}\\
{\large\bf (Generalized Euler Heisenberg Lagrangians)}\\[0.5cm]
{\bf Stephan D\"{u}rr}\\
University of Z\"{u}rich (Switzerland)\\
email : stephan@physik.unizh.ch
\end{center}
\small
The Covariant Derivative Expansion Method (CDEM) is a powerful tool to
calculate oneloop effective actions arising from a heavy particle which is
integrated out in a given gauge field background - for a minimal coupling in
the original theory as well as for a nonminimal Pauli term interaction .

In this talk emphasis is put both on the principles which make this method so
efficient for its specific purpose and on one typical situation where it is a
useful tool inside a more general effective field theory approach to a
phenomenological problem of general interest. It is organized as follows :

In the first section, the idea of Euler Heisenberg Lagrangians is reviewed by
looking at the situation in QED/QCD. It is shown that the virtual photon-photon
/ photon-gluon / gluon-gluon scattering effects due to the box diagram in
lowest order result in an infinite tower of higherdimensional mixed photonic
and gluonic operators which are suppressed by increasing powers of the heavy
particle mass and give the best approximation of the original nonlocal vertex
by a series of local ones. Up to operators of dimension 10 and higher the
additional part in the effective action consists of one dimension 6 (GGG) and
several dimension 8 (FFFF,FFGG,FGGG,GGGG) operators whose coefficients are
finite. For more preliminaries see e.g. ref. [1].

In the second section an exposition of the CDEM as one of the powerful tools
to calculate those (finite) coefficients is given. The CDEM was developped and
promoted in ref. [2] and later improved by several autors [3]. Its main virtue
is the fact that it maintains gauge invariance at every step.

The third section raises the problem of how to determine the phenomenological
impact of additional CP-violation of SM-extensions (left-right-symmetric
models,
multi-higgs-generalizations) generated at $\Lambda\geq m_{t}$ by CP-violating
sunset diagrams (with external gluons attached to it) at the much lower
hadronic scale $\mu\sim m_{s}$. A step-by-step calculation is advocated which
forces one to calculate the effective action stemming from integrating out the
bottom quark coupled to the gluonic background through an additional
CP-violating $\sigma_{\mu\nu}G^{\mu\nu}$ term induced in the previous step.

In the fourth section it is shown how the CDEM can be generalized to this
situation as well. In addition the receipe for including interactions with
an electromagnetic background is given. See ref. [5] for more information.

In the fifth section an RG analysis is performed which turns the present
experimental bound on the neutron electric dipole moment into an upper bound
on the bottom chromoelectric dipole moment at the scale where it arises.
This is intended to be part of the answer to the question wether the previously
mentioned high energy theories are already in conflict with present data .
See ref. [4] for more information.

{\bf References}\\
{}[1] V.A. Novikov etal, Physics Reports C41 (1978) Chap.6 .\\
{}[2] V.A. Novikov, M.A. Shifman, A.I. Vainshtein, V.I. Zakharov,
Sovj. Jour. Nucl. Phys. 39 (1984) 77 or Fortschr. Phys. 32 (1984) 584 .\\
{}[3] L.H. Chan, Phys. Rev. Lett. 57 (1986) 1199 .\\
M.K. Gaillard, Nucl. Phys. B268 (1986) 669 .\\
O. Cheyette, Nucl. Phys. B297 (1988) 183 .\\
{}[4] D. Chang, T. Kephart, W. Keung, T. Yuan, Phys. Rev. Lett. 68 (1992) 439
.\\
S. D\"{u}rr, D. Wyler, in preparation .\\
{}[5] D. Chang, T. Kephart, W. Keung, T. Yuan, Nucl. Phys. B 384 (1992) 147 .\\
S. D\"{u}rr, D. Wyler, in preparation .
\newpage 
\normalsize
\begin{center}
{\large\bf Kaon-nucleon interaction and $\bf \Lambda(1405)$ in dense nuclear
matter}\\[0.5cm]
{\bf T. Waas}, N.\ Kaiser, P.B.\ Siegel and W. Weise\\
Technische Universit\"at M\"unchen, Garching, Germany\\
\end{center}
We examine the free meson-baryon interaction in the strangeness $S=-1$ sector
using
an effective chiral Lagrangian [1]. Potentials are derived from this Lagrangian
and
used in a coupled-channel calculation of the low-energy observables. The
potentials are constructed such that in Born approximation the s-wave
scattering length is the same as that given by the effective chiral Lagrangian,
up to order $q^2$. A comparison is made with the available low-energy hadronic
data of the coupled $\rm K^-p$, $\Sigma\pi$, $\Lambda\pi$ system, which
includes the $\Lambda(1405)$ resonance. Good fits to the experimental data and
estimates of previously unknown Lagrangian parameters are obtained. The
$\Lambda(1405)$ emerges in this approach as a quasi-bound state between an
antikaon and a nucleon.

Including Pauli blocking, Fermi motion and binding of the nucleons we find that
the binding forces between the antikaon and the nucleon are reduced inside
nuclear matter [2]. Therefore the $\Lambda(1405)$ dissolves inside nuclear
matter
at higher densities. Connected with this dynamics of the
$\Lambda(1405)$ is a strong non-linear density dependence of the $\rm K^-p$
scattering amplitude in nuclear matter. The real part of the $\rm K^-p$
scattering
length
changes sign already at a small fraction of nuclear matter density, less
than $0.2\, \rho_0$. This may explain the striking behavior of the
$\rm K^-$-nuclear optical potential found in the analysis of kaonic atom data.

Solving the in-medium kaon dispersion relation [3], we find a strong non-linear
density
dependence of the $\rm K^-$ effective mass and decay width in symmetric nuclear
matter at densities around $0.1$ times normal nuclear matter density $\rho_0$.
At higher
densities the $\rm K^-$ effective mass decreases slowly but stays above
$0.5\, m_{\rm K}$ at least up to densities below $3\,\rho_0$. In neutron matter
the $\rm K^-$ effective mass decreases almost linearly with increasing density
but remains relatively large ($m^*_{\rm K}>
0.65\,m_{\rm K}$) for
$\rho_{\rm n}$ \raisebox{-1.0ex}{$\stackrel{\textstyle<}{\sim}$} $3 \,\rho_0$.
The $\rm K^+$ effective mass turns out to increase very slowly with rising
density.
The different behavior of $\rm K^+$ and $\rm K^-$ effective mass in matter
lead to observable consequences for $\rm K^{\pm}$ production rates in heavy ion
collisions, especially for sub-threshold kinematics. Recent data taken at GSI
are consistent with a lowering of $\rm K^-$ versus $\rm K^+$ in-medium masses
[4].

{\bf References}\\
{}[1] N. Kaiser, P.B. Siegel and W. Weise, Nucl. Phys. A594 (1995) 325.\\
{}[2] T. Waas, N. Kaiser and W. Weise, Phys. Lett. B365 (1996) 12.\\
{}[3] T. Waas, N. Kaiser and W. Weise, Phys. Lett. B (1996), in print.\\
{}[4] R. Barth, E. Grosse, P. Senger et al. (KaoS collaboration), GSI report
10-95, p. 9.
\newpage 
\begin{center}
{\large\bf $\gamma\gamma\to\pi\pi\pi$ and some comments
on $U(1)_A$}\\[0.5cm]
{\bf Johan Bijnens}\\
NORDITA, Blegdamsvej 17\\
DK-2100 Copenhagen \O, Denmark
\end{center}
My talk concentrated on three points:
\begin{enumerate}
\item
The proces $\gamma\gamma\to\pi\pi\pi$ is remarkable in the sense that
the the next-to-leading order is more than order of magnitude larger than the
tree level predictions. Still we expect the next-to-leading order result to be
reasonably accurate.
\item
It has been generally believed that
the presence of the $U(1)_A$ as broken by the anomaly, makes chiral Lagrangians
with a ninth Goldstone Boson rather unpredicitive. Here we show that the
functions of $\phi_0+\theta$ that occur naively can always be reduced by
field redefinitions to just a few constants. As an example, to order $p^2$
the presence of the $\eta'$ introduces at most 4 new constants. This
observation
generalizes to higher order making a proper extension of CHPT to the $U(1)$
sector in principle feasible without recourse to $1/N_c$ counting.
\item
Another problem involving $U(1)_A$ is the fact that in the quenched
approximation the $\eta'$ has a double pole. As a result quenched QCD is
{\em not} a field theory. This leads to counterintuitive results and makes
the chiral limit sick. This part was to make some propaganda for the
work of Bernard, Golterman, Sharpe and collaborators. The topics I discussed
are in [3,4].
\end{enumerate}

{\bf References}\\
{}[1] P.~Talavera et al., hep-ph/9512296, to be
published in Phys. Lett. B, and work in progress.\\
{}[2] J.~Bijnens and M.~Knecht, work in progress.\\
{}[3] C. Bernard and M. Golterman, Phys. Rev. D46 (1992) 853.\\
{}[4] J. Labrenz and S. Sharpe, Nucl. Phys. B(Proc. Suppl.) 34 (1994) 335,
(LATTICE 93)
\end{document}